\newcommand{\HII}{H\,{\tiny II}}

\newcommand{\HI}{H\,{\tiny I}}
\newcommand{\CII}{[C\,{\tiny II}]}

\documentclass{aa}
\usepackage{graphicx}
\usepackage{caption}
\usepackage{txfonts}
\usepackage{xcolor}
\usepackage[colorlinks=true, linkcolor=blue, citecolor=blue, filecolor=blue, urlcolor=blue]{hyperref}

\begin{document} 

\title{Molecular cloud dispersal traced by the ionized carbon 158 $\mu$m line}

\author{L. Bonne\inst{1} 
\and N. Schneider \inst{1}  \fnmsep\thanks{Corresponding author: nschneid@ph1.uni-koeln.de } 
\and S. Dannhauer\inst{1} 
\and E. Keilmann\inst{1} 
\and J.M. Jackson\inst{2} \fnmsep\thanks{Current address: 802/21 Cadigal Ave Pyrmont NSW 2029 Australia}
\and R. Simon \inst{1}
\and A.G.G.M. Tielens \inst{3,4}
\and E. Chambers\inst{5}
\and C. Buchbender\inst{1}
\and J. L. Verbena \inst{1} 
\and S. Kabanovic\inst{6}  
\and T. Faerber \inst{7}
\and L. D. Anderson \inst{7}
\and R. G\"usten \inst{8} 
\and A.M. Jacob \inst{1} 
\and C. Guevara \inst{9}
\and F. Wyrowski \inst{8}
}

\institute{I. Physikalisches Institut, Universität zu K\"{o}ln,  Z\"ulpicher Stra\ss{}e 77, 50937 K\"oln, Germany
\and{Green Bank Observatory, 155 Observatory Road, Green Bank WV 24944 USA}
\and{University of Maryland, Department of Astronomy, College Park, MD 20742-2421, USA} 
\and{Leiden Observatory, PO Box 9513, 2300 RA Leiden, The Netherlands} 
\and{SOFIA Science Center, USRA, NASA Ames Research Center, Moffett Field, CA 94 045, USA}
\and{Astronomisches Rechen-Institut, ZfA Universit\"at Heidelberg, M\"onchhofstraße 12-14, 69120 Heidelberg, Germany} 
\and{Center for Gravitational Waves and Cosmology, West Virginia University, Chestnut Ridge Building, Morgantown, WV26505, USA}
\and{Max-Planck Institut f\"ur Radioastronomie, Auf dem H\"ugel 69, 53121 Bonn, Germany} 
\and{Instituto de Astronom\'ia, Universidad Cat\'olica del Norte, Avenida Angamos 0610, 1270398 Antofagasta, Chile}
}

\date{draft of \today}

\titlerunning{High-velocity wings in CII} 

\authorrunning{L. Bonne}  
 
\abstract
{Feedback from massive stars in the form of radiation and winds impacts the associated host molecular cloud. 
Not only can feedback compress gas and trigger the formation of dense cores that eventually collapse into new stars, it can also disperse cloud material and lead to the destruction of the cloud. Both processes can operate simultaneously, but their relative timescales and the dominant feedback mechanisms remain subjects of active debate. 
Recent observations of the ionized carbon \CII\ 158 $\mu$m line in high-mass star-forming regions have demonstrated that this line is an excellent tracer of the gas dynamics in such environments. Expanding \CII\ shells have been detected, along with high-velocity gas escaping the natal cloud through low-density channels. Motivated by these results, we conducted a systematic analysis of spectrally resolved \CII\ maps obtained with the Stratospheric Observatory for Infrared Astronomy (SOFIA) towards ten high-mass star-forming regions hosting at least one O-type star. This dataset provides a unique perspective on the influence of stellar feedback on the parent molecular cloud. 
Across all regions, we identify high-velocity \CII\ line wings that typically emerge $\sim$5–30~km s$^{-1}$ from the systemic velocity of the host cloud. These velocities exceed the cloud’s escape velocity, indicating that this gas is not gravitationally confined. 
We show that the high-velocity gas exhibits a complex velocity structure and cannot be attributed solely to a single, coherent expanding \CII\ bubble. The amount of material in these erosion flows depends on the evolutionary stage of the molecular cloud and its associated \HII\ region. Once the initial bubble around the cluster ruptures, typically after $\sim$0.1~Myr, gas is expelled from the cloud. 
Estimates of the associated mass ejection rates vary from $\sim$10$^{-3}$~M$_{\odot}$~yr$^{-1}$ for clusters with a single O9V star to 2\,10$^{-2}$~M$_{\odot}$~yr$^{-1}$ for the most massive clusters. The resulting cloud erosion timescales based on these directly observed mass ejection rates typically vary between 2 and 10 Myr after the formation of the first O stars, similar to other indirect measures of molecular cloud lifetimes. These results suggest that stellar feedback is able to remove enough molecular gas to terminate the star formation in the host cloud.
}

\maketitle
\nolinenumbers

\section{Introduction} \label{sec:intro}
Massive (OB) stars inject copious amounts of energy into the interstellar medium (ISM). This can have a significant impact on the molecular clouds out of which these stars have formed: gas compression can trigger further star formation or disperse the parental molecular cloud. 
Concerning the `positive' effect of stellar feedback, two principal frameworks describe the formation of a new generation of stars. In the collect-and-collapse scenario (\citealt{Elmegreen1977}), an expanding \HII\ region accumulates a shell of shocked, cooled neutral gas that subsequently fragments into star-forming clumps. In contrast, the radiation-driven implosion model (\citealt{Lefloch1994, Bertoldi1989}) proposes that feedback acts upon a pre-existing clumpy medium. Whatever the process, many observational studies \citep[e.g.][]{Deharveng2005,Zavagno2006, Fukuda2013} 
show an over-density of stars and star-forming cores in the interface region between the \HII\ region and the molecular cloud, supporting the idea of star formation triggered by stellar feedback.  \citet{Luisi2021} demonstrate that in RCW120, stars can form on short timescales ($<$0.15 Myr) and concluded that positive feedback generally operates on short time periods. However, they point out that the initial cloud structure and turbulent state are shown to strongly determine the impact of radiation and stellar winds on the cloud and its associated star formation. This is further explored in the simulations of \citet{Walch2013}, who investigated radiative feedback and triggered star formation in molecular clouds of varying fractal dimensions. 

The `negative' effect of stellar feedback refers to the erosion of molecular clouds driven by the radiation and stellar winds of massive stars. This process is the focus of the current paper. It is not settled whether feedback truly affects the dense star-forming gas \citep{Watkins2019} and thus can erode the densest clumps in the cloud that are actively forming stars, or if it mostly impacts the inter-clump mass reservoir of the cloud. We note that the formation of stars within a cloud also reduces the total mass reservoir of a molecular cloud and can thus be considered a cloud dispersal process.
Determining whether stellar feedback disperses whole molecular clouds can address the long-standing debate as to whether clouds and their substructures evolve on a crossing timescale \citep[e.g.][]{Elmegreen2000,Hartmann2001,Schneider2023} or if they are close to quasi-static equilibrium \citep{Shu1987, Krumholz2005}. The cloud dispersion timescales also determine whether supernovae occur mostly inside the diffuse or the dense ISM, which affects turbulence injection on galactic scales and galactic outflows \citep[e.g.][]{Rogers2013,Iffrig2017}. In addition, some theories propose that (high-mass) star-forming clouds are in gravitational collapse \citep[e.g.][]{VazquezSemadeni2019}; this idea is gaining support from observations \citep[e.g.][]{Schneider2010,Peretto2013,Motte2018}. Stellar feedback is thus instrumental in limiting the star formation rates to the low observed values \citep{Evans2009,Lada2010}.

To trace the influence of stellar feedback on molecular clouds, it is particularly important to probe photodissociation regions (PDRs) as they form on the surface layers of clumps within molecular clouds where far-ultraviolet (FUV) radiation (E$_{photon}$ = 6-13.6 eV)  from massive stars regulates the heating and chemistry \citep{Tielens1985,Hollenbach1999,Wolfire2022}. The fine structure line of ionized carbon at 158 $\mu$m (\CII) is a well-established tracer of PDRs, and spectrally and spatially resolved \CII\ observations unveil the dynamics at the interface with molecular clumps.
Over the last few years, high-spectral-resolution (R~$>$~1~000~000) data cubes of the \CII\ fine structure line have become available towards massive star-forming regions with the development of the (up)GREAT receiver \citep{Heyminck2012,Risacher2016,Risacher2018} deployed on board the Stratospheric Observatory for Infrared Astronomy (SOFIA; \citealt{Young2012}). These data provide a unique view into the PDR dynamics and how stellar feedback affects molecular cloud evolution. Early \CII\ observations with SOFIA  of PDRs surrounding O stars  demonstrated the presence of high-velocity gas (with velocities up to 20-25 km s$^{-1}$ relative to the systemic velocity of the host molecular cloud), which is visible in the form of broad spectral wings \citep{Simon2012,Schneider2018} that are blue- or redshifted, or both. These \CII\ high-velocity wings can also be found when inspecting the spectra observed with the \textit{Herschel} HIFI detector \citep[e.g.][]{Pineda2013,Goicoechea2015}. Interest in high-velocity \CII\ emission gained traction with the first SOFIA results from the Orion~A map \citep{Higgins2021} and the FEEDBACK legacy survey \citep{Schneider2020}. Such observations of M42 in Orion \citep{Pabst2019} and RCW~120 \citep{Luisi2021}  revealed that the \CII\ emission has an expanding spherical shell morphology with velocities of up to $\sim$15 km s$^{-1}$. In both studies, this geometry was interpreted as a classical expanding bubble, similar to \HII\ region expansion, described by \citet{Spitzer1978} as a spherical ionized region around a hot star that grows over time as the high internal gas pressure drives a shock front outwards into the surrounding neutral medium. \citet{Weaver1977}, in addition, considered stellar winds to be a driving mechanism for expanding \HII\ regions. The occurrence of high-velocity expansion in the PDR has important implications for the evolutionary timescales in these regions as the estimated bubble lifetimes for both regions are only $\sim$0.1 Myr \citep{Pabst2019,Luisi2021,Faerber2025,Dannhauer2025}.

Further work with the FEEDBACK legacy survey and other independent observing programmes carried out with SOFIA found that expanding \CII\ shells (reaching expansion velocities of at least 10 km s$^{-1}$) appear to be present in many \HII\ regions around OB stars \citep[e.g.][]{Tiwari2021,Beuther2022,Bonne2022b,Tram2023}. A first statistical study of Galactic \HII\ regions associated with \CII\ emission by \citet{Faerber2025} found at least partial expanding \CII\ shell morphologies with velocities $\ge$10~km~s$^{-1}$ in 34$\%$ of their sample of 35 sources. This yields similar dynamical timescales of $\sim$0.1 Myr for all identified expanding shell candidates. Recently, \citet{Dannhauer2025} discovered a late stage of an expanding \CII\ bubble that only consists of a slowly expanding (v$<$2 km s$^{-1}$) ring without a 3D spherical structure. Dedicated hydrodynamic simulations of an expanding \HII\ region with a central B0.5 star, placed within a flat molecular cloud, reproduced these observations. It was shown that a bubble-like spherical expansion happens only in the first 0.1 Myr. As pointed out in \citet{Faerber2025}, these short timescales raise an important tension as O star clusters typically reach ages of up to $\sim$3-5 Myr before their first supernova.

\citet{Bonne2023c} reported the presence of a rapidly expanding \CII\ shell in RCW79, a result later confirmed by \citet{Faerber2025}. However, a detailed analysis of the \CII\ emission in \citet{Bonne2023c}, employing clustering techniques and expanding-bubble models, demonstrated that the kinematic structure of this region is considerably more complex. RCW79 is best explained by a combination of a fragmented expanding \CII\ shell and mostly neutral gas that is ejected from the cloud through low-density holes. 
This idea of \CII\ gas ejection was also considered in a few other recent works \citep{Beuther2022, Bonne2022b,Kabanovic2022,Jackson2024}.

In this context, it is essential to distinguish between the concepts of `(bubble) expansion' and `mass-ejection'. 
Expansion corresponds to any coherent 3D outward motion of a gas layer from a centre of expansion, which is typically a single OB star or an ionizing OB star cluster. This expansion is commonly identified by elliptical features in position-velocity (PV) diagrams of \CII\ emission \citep{Pabst2019,Luisi2021}. Importantly, during the expansion, the outward moving gas layer can remain gravitationally bound to the host cloud. In a homogeneous, non-turbulent medium, the velocity of such an expansion can reach the sound speed in the \HII\ region of $\sim$10 km s$^{-1}$ for thermally driven flows. However, even higher velocities have been observed (see the references above) and are commonly attributed to energy-driven expansion resulting from stellar winds from massive stars. Mass ejection, on the other hand, is defined as any parcel of gas that is moving in an outward direction with respect to its host molecular cloud and will eventually leave that cloud to become gravitationally unbound and join the diffuse ISM. For this to happen, the observed parcel of gas has to move at a relative velocity above the escape velocity of the host molecular cloud, which is typically of the order of 5 km~s$^{-1}$ \citep[e.g.][]{Kim2018}. Unlike expansion, mass ejection does not require a coherent 3D structure and can involve individual high-velocity components and substructures.
This process can include neutral gas that is sheared off and entrained by photo-ionized or collisionally ionized plasma streaming along the surface of the opening cones of the bubble.

Note that expansion and mass-ejection are not mutually exclusive. Expanding shells with velocities above the escape velocity can be part of the overall mass-ejection process in a region \citep[e.g.][]{Walch2012, Bonne2023c}. 
In this context, two additional observational aspects of expanding-shell candidates are noteworthy. First, 
in some regions presented by \citet{Faerber2025}, the expanding shell morphology is only observed in parts of the region and does not cover the full region as expected by an expanding bubble. Second, although 66\% of sources in \citet{Faerber2025} display no evident expanding shell morphology, the \CII\ spectra show that these regions have high-velocity \CII\ emission reaching 20–25~km s$^{-1}$.

In this paper, we build on these results and explore the properties of observed high-velocity \CII\ gas in a cloud sample observed with SOFIA (Sect.~\ref{sec:obs}) to propose a more comprehensive scenario. First, we demonstrate the ubiquity of \CII\ high-velocity gas, compare the \CII\ emission distribution with the molecular cloud structure revealed by CO observations, and calculate the associated dynamical timescales (Sect.~\ref{sec:results}). We then (Sect.~\ref{sec:discuss}) address the implications of this ubiquitous \CII\ high-velocity gas, quantify the associated mass ejection rates in molecular clouds, and discuss what this implies for molecular cloud lifetimes. The paper is summarized in Sect.~\ref{sec:conclusions}. 

\begin{table*} []
    \centering
    \small
    \caption{Cloud properties.}
    \begin{tabular}{lccccccccccc}
    \hline
    \hline
  Source   & D                   & M$_{cloud}$ & r$_{cloud}$ & O-stars$^+$ & t$_{cluster}$       & v$_{cloud}$ & v$_{b}$ & v$_{r}$ & v$_{esc}$ \\
           & (kpc) & (10$^3$ M$_\odot$)    & (pc)   &     & (Myr)     & (km s$^{-1}$)  & (km s$^{-1}$)  & (km s$^{-1}$)  & (km s$^{-1}$)  \\
 \hline
  M42      & 0.414$\pm$0.007 (1) & 2.45    & 2.4    & 1 O7V & 2.5$\pm$0.5 (11)    &     10.1       & [-11,7.1]   & [13.1,17.1] & 3.0   \\
  M16      & 1.706$\pm$0.007 (2) & 87.1    & 18.3   & 1 O4,$\sim$10 late O & 1.3$\pm$0.2 (2)     &      21.8      & [3,-15.4]   & [30,41]     & 6.4  \\
  M17      & 1.9$\pm$0.1     (3) & 483.8(*)   & 28.1   & 2 O4,$\sim$10 late O & 0.58$\pm$0.15 (12)  &        19.8    & [-11,7.7]   & [37,52]     & 12.1 \\
  RCW120  & 1.68$^{+0.13}_{-0.11}$ (4)& 10.5 & 4.5 & 1 O8 & 1.2$^{+1.8}_{-1.2}$ (13)(**) &  -7.5  & [-27,-12]   & [-3,18]     & 4.5  \\
  RCW79   & 3.9$\pm$0.4     (5) & 17.4    & 10.1   & 2 O4,$\sim$10 late O & 2.3$\pm$0.5 (13)(**)  &     -47.2      & [-70,-51.1] & [-38,-26]   & 3.9 \\
  RCW36   & 0.95$\pm$0.05   (6) & 15.2    &  6.5   & 2 O9 & 1.1$\pm$0.6 (14)     &      6.6     & [-12,2.2]   & [11.2,22]   & 4.5  \\
  NGC~7538 & 2.65$^{+0.12}_{-0.11}$ (7)& 130.4 & 22.0 & 1 O3, 1 O9 & 2.2$\pm$0.9 (15)   &     -55.8      & [-80,-62.9] & [-48.7,-33] & 7.1  \\
  RCW49   & 4.16$\pm$0.33 (8)   & 60.9    & 16.9   & 2 WR,12 early O & 1.25$\pm$0.75 (8)    &      1.5     & [-30,-10]   & [20,30]     & 5.6  \\
  W40      & 0.502$\pm$0.004 (9) & 16.8    & 7.4    & 1 O9.5 & 0.9$\pm$0.9 (9)      &      6.0     & [-14,1.6]   & [10.4,23]   & 4.4  \\
  S106     & 1.3$\pm$0.1 (10)    & 4.45    & 3.0    & 1 O9 & 0.1(16)(***)        &       -0.5    & [-14,-4.2]    & [3,15]      & 3.6  \\
 \hline
    \end{tabular} \\
   \tablefoot{Columns 1 and 2 give the source name and its distance. Almost all distance estimates are based on trigonometric parallax measurements except for RCW49 where the distance is based on photometric estimates. 
    Column 3 is the mass determined from {\it Herschel} dust column density, most of the values are reported in \citet{Schneider2022}. Column 4 is the equivalent radius ${\rm r}_{cloud}=\sqrt{A/\pi}$, given by the area $A$ of the cloud. The area is defined by the A$_v$ = 1 contour \citep{Schneider2022} or taken from the literature for M42 and S106. Column 5 lists the O-star content of the region, partly taken from \citet{Schneider2020}, and updated literature values. Column 6 specifies the estimated age of the cluster. Column 7 is the line centre velocity of the molecular cloud, determined from a Gaussian fit to the $^{13}$CO line. Columns 8 and 9 specify the blue ($b$) and red ($r$) range of high-velocity emission. Column 10 gives the escape velocity. }
    \tablebib{
    (1) \citet{Menten2007}, 
    (2) \citet{Stoop2023}, 
    (3) \citet{Wu2019}, 
    (4) \citet{Kuhn2019},
    (5) \citet{Bonne2023c}, 
    (6) \citet{Zucker2020}, 
    (7) \citet{Moscadelli2009},
    (8) \citet{Zeidler2015}, 
    (9) \citet{Comeron2022},  
    (10)\citet{Xu2013},  
    (11) \citet{DaRio2010}, 
    (12) \citet{Hanson1997}, 
    (13) \citet{Martins2010}, 
    (14) \citet{Ellerbroek2013}, 
    (15) \citet{Puga2010},
    (16) \citet{Comeron2018}. \\
    $^+$ If known, the spectral type of the O-star is given. WR signifies Wolf-Rayet star. 
    (*) The mass is an upper limit from \citet{Schneider2022}, values in the literature are 3.2 10$^5$ M$_\odot$ from dust \citep{Dupac2002} and range, based on CO observations, from 1.35\,10$^5$ M$_\odot$ \citep{Povich2009} to 2.8\,10$^5$ M$_\odot$ \citep{Elmegreen1979} 
    (**) The results for RCW120 and RCW79 rely solely on age estimates of the ionizing stars, not on cluster-member ages.
    (***) The age estimate of the central O9 star of S106 is not well constrained.}
    \label{tab:tabObsSummary}
\end{table*}

\section{Observations} \label{sec:obs}

\subsection{SOFIA \CII\ data}
For this study we made use of the \CII\ spectral data cubes from the FEEDBACK legacy survey (program\_ID: 07\_0077, PIs A. Tielens and N. Schneider), the Orion legacy programme (program\_ID: 04\_0066, PI A. Tielens), and a S106 Guaranteed Time programme \citep{Simon2012}, obtained with the (up)GREAT receiver on board SOFIA \citep{Heyminck2012,Risacher2016,Risacher2018}. From the FEEDBACK programme, we used data from the high-mass star-forming regions RCW49, NGC7538, RCW36, RCW79, RCW120, W40, M17, and M16 (Table \ref{tab:tabObsSummary}). With this selection, we focused on the regions that contain one or more O stars with an available age estimate for the cluster in the literature. The observational details and data reduction strategy for the FEEDBACK and Orion mapping programmes are presented in \citet{Schneider2020} and \citet{Higgins2021}, respectively.

The SOFIA observations were all performed in the on-the-fly OTF mode and calibrated with the (up)GREAT pipeline. 
The intrinsic spatial resolution at 158 $\mu$m is 14.1$^{\prime\prime}$, but to achieve a better signal-to-noise ratio (S/N), we convolved the data to spatial resolution of 20$^{\prime\prime}$ and used a grid of 8$^{\prime\prime}$. We employed data cubes on a main beam brightness temperature scale with a velocity resolution of 0.5 km s$^{-1}$ with a resulting noise rms between 0.48 K and 0.58 K per spectral bin of 0.5 km s$^{-1}$. All data cubes were reduced with a dedicated principal component analysis (PCA) technique developed by C. Buchbender (priv. comm.) for SOFIA \CII\ data. PCA was applied for all sources in the FEEDBACK programme, as it strongly reduces striping in the data \citep{Tiwari2021, Kabanovic2022, Schneider2023}. For the red-green-blue (RGB) plots, we made use of the masked-moment map procedure, which is a weighting technique, to produce high-quality line-integrated maps (zeroth-moment; \citealt{Adler1992}). 

\subsection{Complementary CO and dust data}
We complemented the \CII\ observations with $^{12}$CO and $^{13}$CO data cubes of the associated molecular clouds, which were used for spectral-line comparisons and, in the case of S106, for mass estimates. For the FEEDBACK sources, $^{12}$CO(3-2) and $^{13}$CO(3–2) (lines at 345.796 and 330.588 GHz) maps were obtained with the APEX 12~m sub-millimetre  telescope \citep{Guesten2006} for targets in the southern hemisphere, while $^{12}$CO(2-1) and $^{13}$CO(2–1) (lines at 230.538 and 220.399 GHz) maps were acquired with the IRAM 30m telescope for NGC7358 \citep{Beuther2022} and M42 \citep{Goicoechea2020} in the northern hemisphere. For S106 we only used the $^{13}$CO(1-0) large map of the molecular cloud published in \citet{Schneider2007}. The final CO data cubes have a uniform spatial resolution of $\sim$20$^{\prime\prime}$, matched to that of the \CII\ observations.

Molecular cloud mass and radius are mostly determined from {\it  Herschel} column density maps. For M16, M17, NGC7538, we used the values given in Table 2 in \citet{Schneider2022}, based on data from the HOBYS key programme \citep{Motte2010}. For RCW36, RCW120, RCW49,  RCW79, and W40, we employed the column density maps performed with PPMAP \citep{Marsh2017}, making use of the Hi-GAL survey \citep{Molinari2010,Molinari2016}. For M42, the column density map published in \citet{Takemura2023}, based on the data from the Gould Belt survey \citep{Andre2010}, was used. The maps have angular resolutions of 12$''$ (PPMAP) and 18$''$ (HOBYS and Gould Belt). For the mass and radius determination, we chose a lower limit of visual extinction\footnote{The molecular hydrogen column density and visual extinction are related by N(H$_2$) = A$_v$ 0.94 10$^{21}$ cm$^{-2}$ mag$^{-1}$.} of A$_v$ = 1~mag as the border of the molecular cloud (see \citealt{Schneider2022} for a justification based on column density probability distribution functions). For S106, we adopted the mass estimate from FCRAO $^{13}$CO(1–0) (line at 115.271 GHz) data published in \citet{Schneider2007}. Distance, mass, and radius, together with estimates of star (cluster) ages for each region are summarized in Table \ref{tab:tabObsSummary}.

\begin{figure}
    \centering
    \includegraphics[width=\hsize]{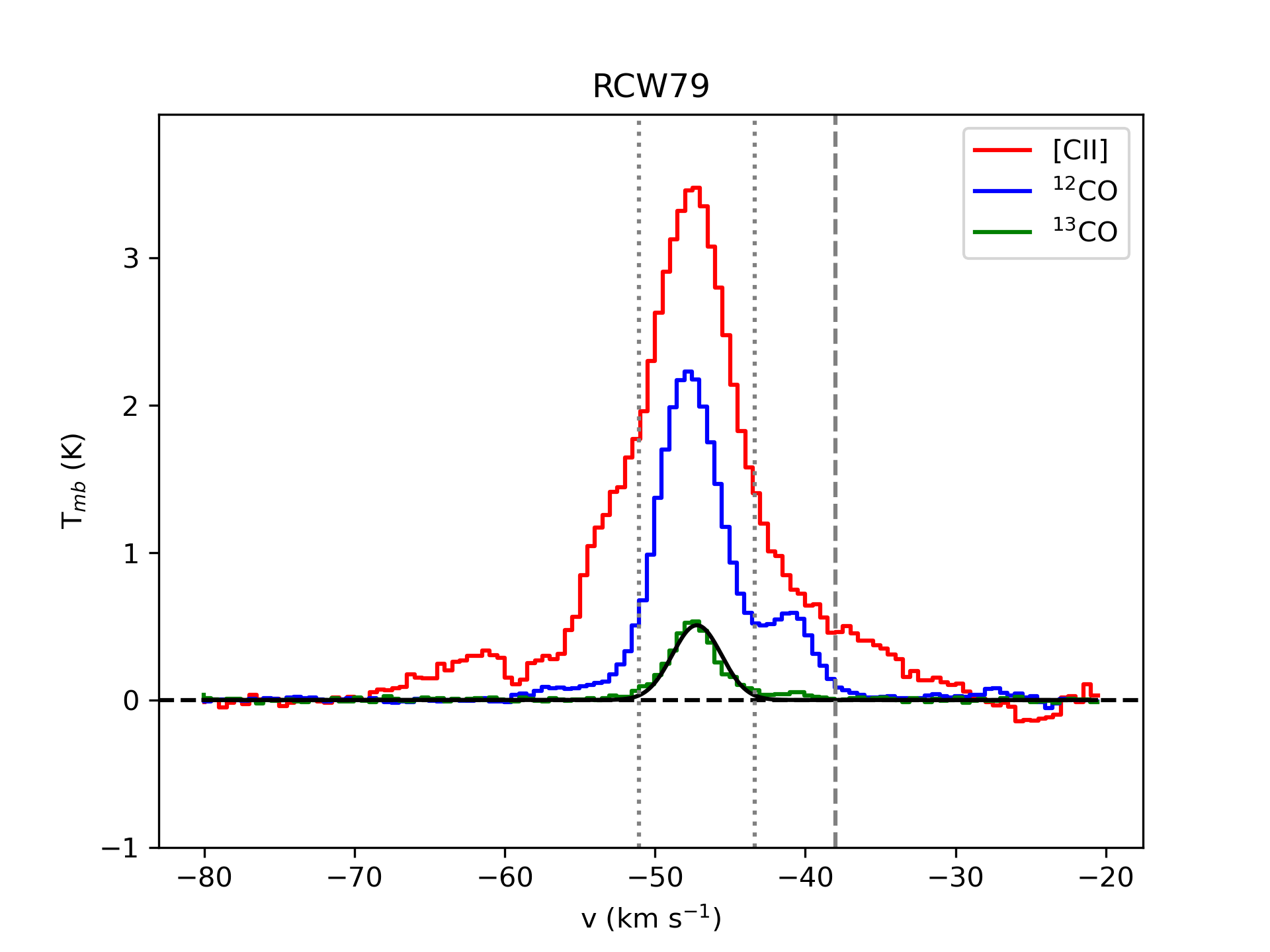}
    \caption{Averaged \CII\ and $^{12}$CO and $^{13}$CO(3-2) spectra for RCW79. The spectra are averaged over the full region mapped in \CII. The vertical dotted lines mark the velocity intervals associated with the high-velocity wings (defined by the escape velocity), while the vertical dashed line indicates the revised threshold for the onset of high-velocity emission, chosen to exclude contamination from fore- or background sources. The Gaussian fit to the $^{13}$CO line is in black.
    }
    \label{fig:avSpecM17-RCW79}
\end{figure}

\begin{figure*}
    \centering
    \includegraphics[width=0.98\hsize]{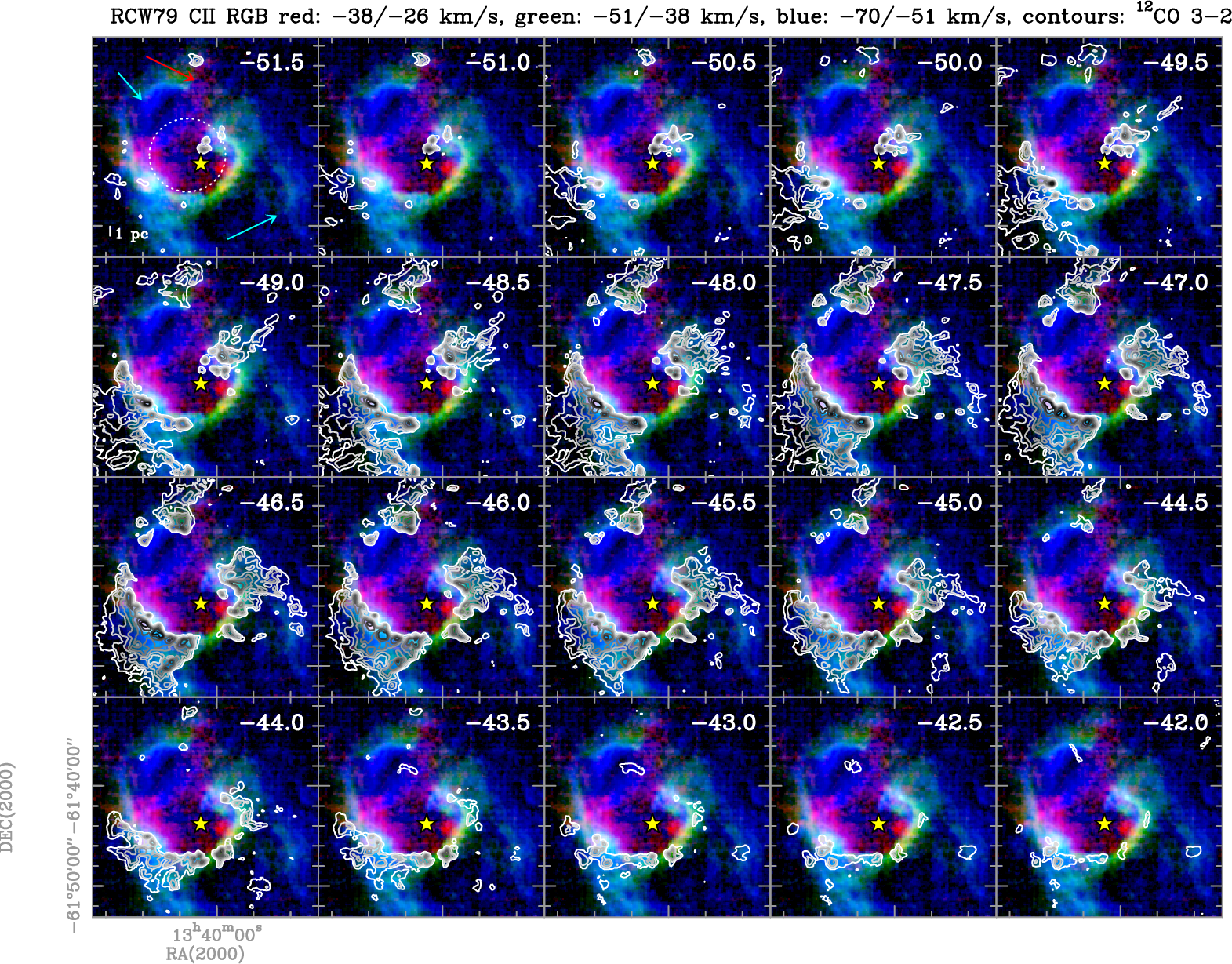}
   \caption{Combined RGB plot in \CII\ and CO channel map of RCW79. Each panel displays the distinct velocity ranges (blue and red for the high-velocity blue- and redshifted emission, green for the bulk emission) of \CII\ as an RGB plot with contours (3 to 25.4 K km s$^{-1}$ by 3.2 K km s$^{-1}$) of $^{12}$CO(3-2) emission overlaid. The corresponding CO velocity is given in the upper-right corner. In the first panel, the expanding \CII\ bubble is approximated by a dotted circle. The exciting star cluster is indicated with a yellow star. Examples of high-velocity blue- and redshifted \CII\ gas not associated with the expanding bubble are indicated with cyan and red arrows, respectively.}
    \label{fig:chanmap-rcw79}
\end{figure*}

\section{Results} \label{sec:results}
\subsection{Average \CII\ spectra of the regions}
Figure~\ref{fig:avSpecM17-RCW79} presents as an example the averaged \CII\ and CO spectra for RCW79. Spectra for all other sources are provided in Appendix \ref{sec:avSpec}. The spectral profiles exhibit significant variation among regions: some (e.g. RCW79 and M42) display a single central velocity component, whereas others (e.g. M17 and RCW49) reveal up to three distinct velocity components. Except for W40 and RCW36, the line shape in these average spectra is not too much affected by the frequently observed \CII\ self-absorption \citep{Guevara2020,Kabanovic2022} that could mimic several line components. However, in each \CII\ spectrum we find that the brightest component also corresponds to peak emission in the $^{12}$CO and $^{13}$CO lines, and can thus be regarded as the bulk emission of the PDR and the associated molecular cloud. The velocity v$_{cloud}$ corresponding to the bulk emission was determined by a Gaussian fit to the $^{13}$CO line. A remarkable feature in the \CII\ spectra for all regions are high-velocity emission wings outside of the bulk velocity without a CO counterpart. This \CII\ emission reaches velocities of up to several 10 km s$^{-1}$ relative to the central velocity of the host molecular cloud. We defined here the high-velocity ranges as the spectral bins that are outside of the escape velocity (see Sect.~\ref{subsec:esc}).

We note that high-velocity \CII\ emission was also found in all other regions observed in \CII\ with identified O stars that are not included in this study because they have no published cluster age, such as S235 (\citealp{Anderson2019}) 30 Doradus (\citealp{Tram2023}), and Nessie A (\citealp{Jackson2024}). This high-velocity \CII\ emission is thus ubiquitous in regions surrounding O star clusters.

\subsection{Escape velocities for the clouds} \label{subsec:esc}
The escape velocity is a measure of the gravitational binding energy and to first order determines whether internal motions or stellar feedback can remove material from the cloud or whether the gas remains confined. A cloud with a high v$_{esc}$ is difficult to disperse,
while  a cloud with low v$_{esc}$ is easily cleared by stellar radiation and wind. The gravitational escape velocity is given by  
$ v_{esc} = \sqrt{2 \,  G \,\text{M}_{cloud}/\text{r}_{cloud}}$,   
where $G$ is the gravitational constant and M$_{cloud}$ and r$_{cloud}$ are the mass and radius of the cloud, respectively.
To define the velocity interval confined by the escape velocities, we used the central velocity of the molecular cloud, estimated from a single Gaussian fit to the average $^{13}$CO spectrum (indicated in Figs.~\ref{fig:avSpecM17-RCW79} and \ref{fig:avSpecs}). As outlined above, this velocity component corresponds also to the \CII\ bulk emission. The resulting interval with velocities below the escape velocity is then defined as v$_{cloud}$ $\pm$ v$_{esc}$. The centre velocity and the ranges of high-velocity emission are given in Table~\ref{tab:tabObsSummary} together with the mass, derived mostly from {\it Herschel} column density maps. We note that here we used the current mass of the cloud, not trying to extrapolate the original cloud mass as was done in \citet{Bonne2023c}. In addition, in this simplified approach we ignored magnetic fields that help support a cloud against gravity and could slow down the high-velocity gas, thus requiring higher velocities than the gravitational escape velocity alone to get out of the molecular cloud. Note also that because $v_{esc}$ is a function of radius, in the case of constant density, gas near the centre feels a deeper potential than gas near the edge and can still be bound. On the other hand, if the density profile follows a singular isothermal sphere, the escape velocity is roughly constant with radius. 

\begin{figure}
    \centering
    \includegraphics[width = 0.85\hsize]{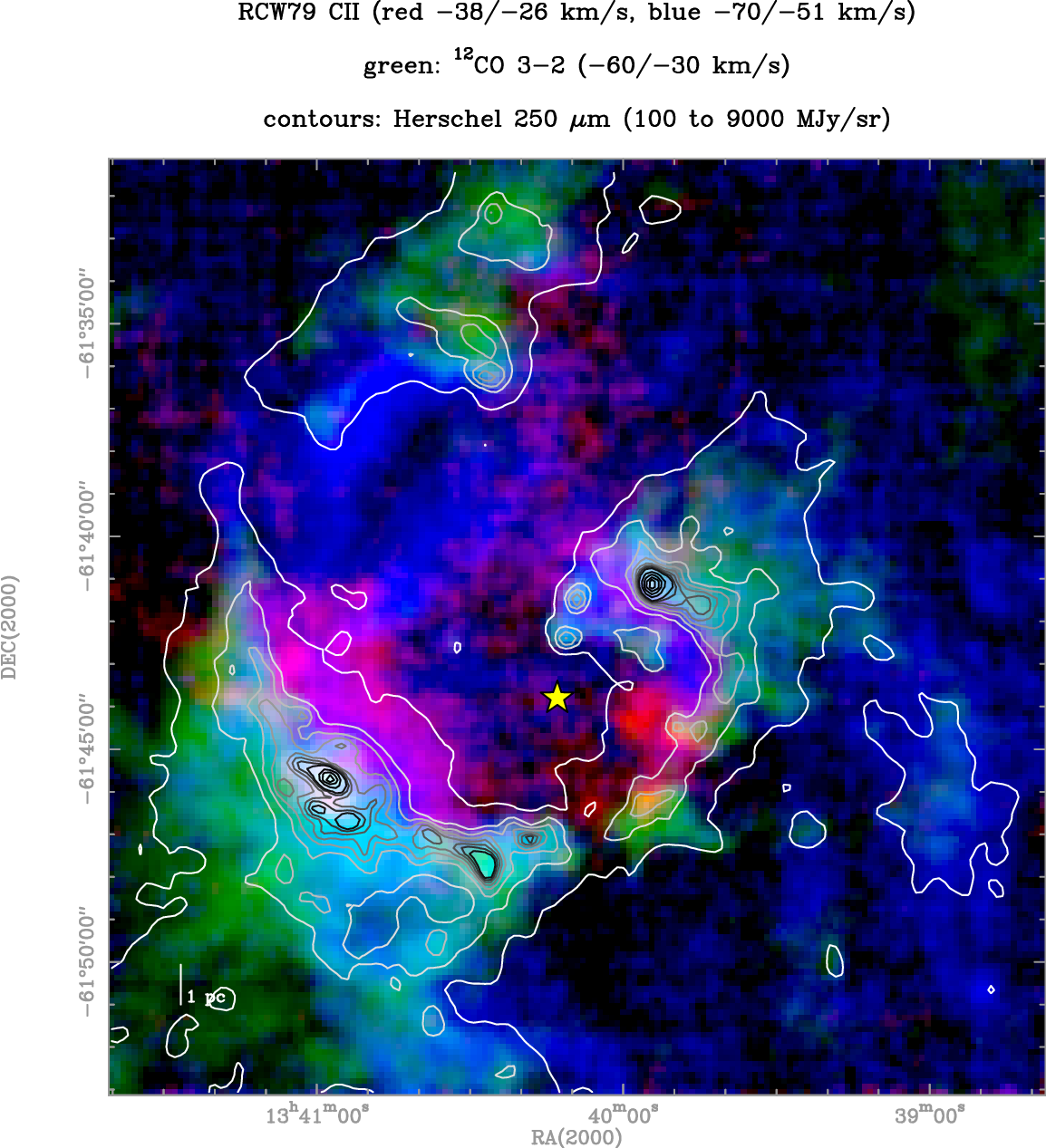}
    \caption{RGB plot with dust overlay of RCW79. The image shows \CII\ high-velocity emission in red and blue, and CO emission over the full velocity range in green. Overlaid are contours of {\it Herschel} 250 $\mu$m emission. The yellow star indicates the exciting cluster of RCW79.}
    \label{fig:rgb-dust}
\end{figure}

\begin{figure}
    \centering
    \includegraphics[width=0.85\hsize]{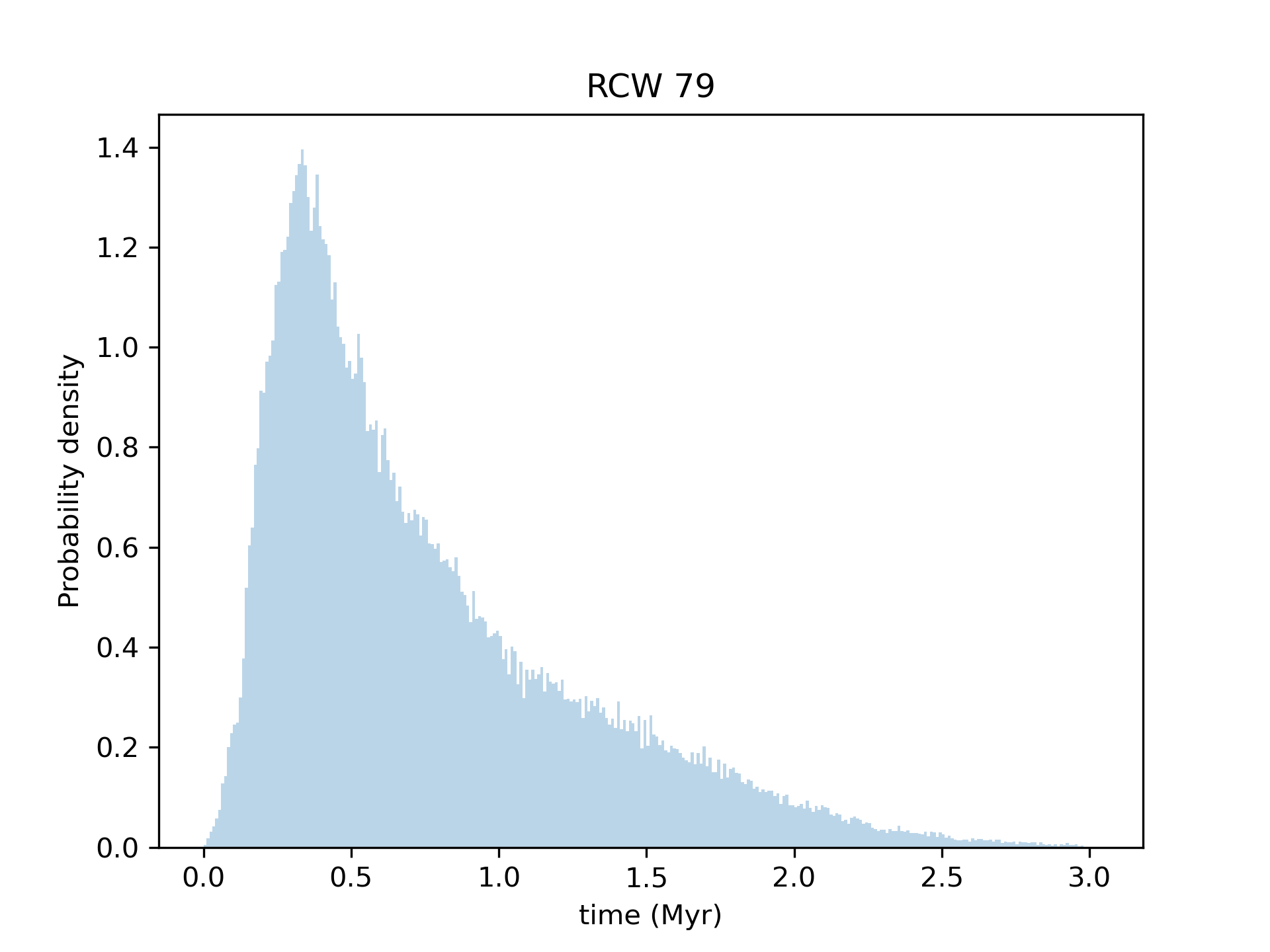}
    \caption{Distribution for the dynamical timescales associated with the high-velocity wings in RCW79. The other studied star-forming regions have similar distributions.}
    \label{fig:dyn_age_hists}
\end{figure}

\subsection{Spatial distribution of the high-velocity gas} \label{subsec:spatial}
The spatial distribution of blue- and redshifted \CII\ high-velocity gas with respect to the associated molecular cloud is illustrated in Fig.~\ref{fig:chanmap-rcw79} for RCW79, with corresponding maps for the remaining regions presented in Appendix~\ref{sec:intMaps}. Position–velocity diagrams that also help illustrate the velocity structure for most sources are shown in \citet{Faerber2025}. Overall, the morphology of the high-velocity \CII\ gas is complex and varies significantly among regions.
Despite this diversity, several systematic trends emerge. In many cases, a part of the high-velocity emission traces the well-known shell structures, with red- and blueshifted components\footnote{The higher abundance of blue- vs redshifted \CII\ shells is probably a selection effect as observational studies focused on optically visible \HII\ regions; i.e. located on the front of the cloud facing the observer \citep{Faerber2025}.} located interior to the bright PDR emission (shown in green). Prominent examples include RCW49, RCW79, RCW120, and NGC7538. In contrast, the bipolar nebulae RCW36, W40, and S106 do not exhibit clear expanding-shell signatures. M16 and M17 display complex emission patterns that hinder the identification of coherent bubbles, although M17 is classified as a source with an expanding \CII\ shell by \citet{Faerber2025}. M42 displays a circular spatial morphology for the blueshifted high-velocity gas. 

Particularly noteworthy is the high-velocity emission extending beyond the bright PDR structures, for example the ring-like features indicated in green in RCW120 (Fig.~\ref{fig:rgb2}) and RCW79 (Fig.~\ref{fig:chanmap-rcw79}) that enclose the \HII\ region (the  bright PDRs are surface features located on dense molecular clumps). The CO channel maps (contours in Figs.~\ref{fig:chanmap-rcw79} and~\ref{fig:rgb1}--\ref{fig:rgb5}) illustrate how fragmented the CO clouds are, spatially and kinematically. Several openings are visible for RCW79 in various velocity bins and it is through these holes that the high-velocity gas could escape. This was also shown for RCW20 in \citet{Anderson2015}. The high-velocity \CII\ emission is then possibly a mixture of accelerated inter-clump gas and eroding or entrained gas from the dense PDR surfaces. This would support classical models such as those from \citet{Matzner2002}, promoting that ionization-driven champagne flows should be an efficient mechanism for disrupting molecular clouds.
Prominent redshifted emission outside the PDR and molecular cloud is observed in the eastern parts of NGC7538 and RCW36, as well as in the southern region of W40. In M16, M17, RCW49, and M42, this component appears as spatially patchy emission. Blueshifted high-velocity emission not associated with expanding \CII\ shells is generally less prominent and, in most cases, remains closely connected in the plane-of-the-sky to the bright PDR emission at the systemic velocity (e.g. RCW79, RCW120, NGC7538, and M17). In contrast, in S106, W40, and RCW36, the blueshifted component is located further from the bulk emission, while in M16 and M17 it is distributed unevenly. 

Here, and for the calculations in Sect.~\ref{subsec:mass-ejection}, we assumed that the high-velocity \CII\ emission is arising from the neutral, atomic gas. This is a justified assumption since it was shown that the gas in the bubble shell and  the entrained gas is typically atomic with a low density of around n$_{\rm H} \sim$10$^2$ to a few times 10$^3$ cm$^{-3}$ \citep[e.g.][]{Luisi2021, Bonne2023c}. Some \CII\ emission can stem from the ionized gas of the \HII\ region. \citet{Luisi2021} deduced for RCW120 a maximum of 20\% of the observed \CII\ can arise in the \HII\ region and \citet{Bonne2023c} found that the percentage is up to 30\% of the high-velocity wings. However, the \CII\ emission we associated with erosion is typically spatially distinct from the location of the \HII\ region and located mostly beyond the \HII\ region and molecular cloud. 
Another possibility is that \CII\ arises from CO-dark, molecular gas. These areas are difficult to trace. In Cygnus X, \citet{Schneider2023} showed by comparing CO, \CII\ and \HI\ self-absorption that there is indeed a reservoir of CO-dark molecular gas that actually builds up denser molecular clouds. We here produced an overlay between high-velocity \CII\ emission, CO emission in the full velocity range, and dust emission from {\it Herschel} at 250 $\mu$m to possibly trace CO-dark gas for RCW79 (Fig.~\ref{fig:rgb-dust}). The 100 MJy/sr lower contour is somewhat arbitrary, but traces weak 250 $\mu$m emission associated with cool to warm dust. There are indeed some areas where we observe high-velocity \CII\ emission that correlates with dust, but not with CO (e.g. around RA(2000)=13$^h$41$^m$30$^s$, Dec(2000)=$-$61$^\circ$40$'$). This could be CO-dark molecular gas, but since dust traces the atomic and molecular phase, we cannot attribute these regions fully to molecular gas. We performed this exercise for all sources and found that areas with dust and \CII\ emission, but no CO, contribute to roughly 10-20\% of the high-velocity \CII\ emission distribution. 

For completeness, we note that a possible source of high-velocity \CII\ emission are protostellar outflows associated with young stellar objects (YSOs). \citet{Kavak2022a, Kavak2022b} identified several very localized high velocity \CII\ emission features that were ascribed to bullets driven by low-mass protostellar jets that have pierced the Orion Bubble wall.
However, at the locations of such sources (e.g. in RCW120 and RCW79 there are isolated YSOs within the ring structure), we do not observe a notable increase in blue- and redshifted high-velocity \CII\ emission. In addition, if most of the emission would be associated with YSO-driven outflows the red- and blueshifted high-velocity gas should display clear bipolar (or rarely unipolar) structures centred around the YSOs, which is not observed. In addition, protostellar outflows do not extend over several parsecs as it is the case in our picture of erosion flows. 

As a first qualitative interpretation, the high-velocity emission not associated with well-defined expanding \CII\ shells likely traces gas entrained from PDR surfaces and channelled through low-density pathways within an inhomogeneous molecular cloud. Stellar winds and radiation drive the dynamics, overcoming gravity and allowing gas to escape the cloud. This process gradually erodes the molecular cloud by removing material over time and should depend on factors such as the density and volume-filling factor of molecular clumps, as well as the strength of radiative and wind feedback. We return to this issue in Sect.~\ref{sec:discuss}.

\subsection{A representative dynamical timescale for the high-velocity gas} \label{subsec:time}

To better understand the nature of the high-velocity gas, we started by calculating the associated dynamical timescale, t$_{dyn}$ = d/v$_{high}$, where d is the distance that a high-velocity gas parcel has travelled from where it got accelerated by the impact of stellar feedback and v$_{high}$ the relative velocity with respect to the cloud. Assuming there is no preferred direction in 3D space for the high velocity gas and spherical geometry, we can estimate t$_{dyn}$ from the high-velocity wings and distances in the plane-of-the-sky. The assumption that the plane-of-the-sky distance is similar to the line-of-sight (LOS) distance is supported by studies of high-velocity gas in, for example, RCW36 \citep{Minier2013,Bonne2022b} and estimates of the FUV field strength near the high-velocity gas in RCW79 using PDR models \citep{Bonne2023c}. The ratio of the FUV field strength over the total luminosity of the ionizing cluster is directly related to the distance.

To estimate a t$_{dyn}$ for each region, we calculated an intensity-weighted timescale averaged over all pixels. First, we identified all spatial pixels that have high-velocity wings detected at the 3$\sigma$ level. For each of the spatial pixels we calculated the distance from the centre of the O star cluster. Correcting this distance for each pixel by a factor of$\sqrt{2}$, to convert the 2D distance in the plane-of-sky to a 1D estimate, provides us a range of values for the distances (d) in the LOS. For all detected (T$_{mb}$ $>$ 3$\sigma$) high-velocity spectral bins in these pixels, we used their intensity and velocity relative to the velocity of the bulk emission of the cloud. A single representative value for the high-velocity gas in each region is then calculated using
\begin{equation}
    v_{high} = \Sigma_{pix}\left( \Sigma_{i}\frac{T_{mb, pix, i}\cdot \|v_{i} - v_{cloud}\|}{T_{mb, pix, i}} \right)
,\end{equation}
where T$_{mb, pix, i}$ is the main beam brightness of the spectral bin (i) in a given pixel, v$_{i}$ the local standard of rest (LSR) velocity of the spectral bin and v$_{cloud}$ the LSR velocity of the bulk molecular cloud.

\begin{figure}
    \centering
    \includegraphics[width=0.95\hsize]{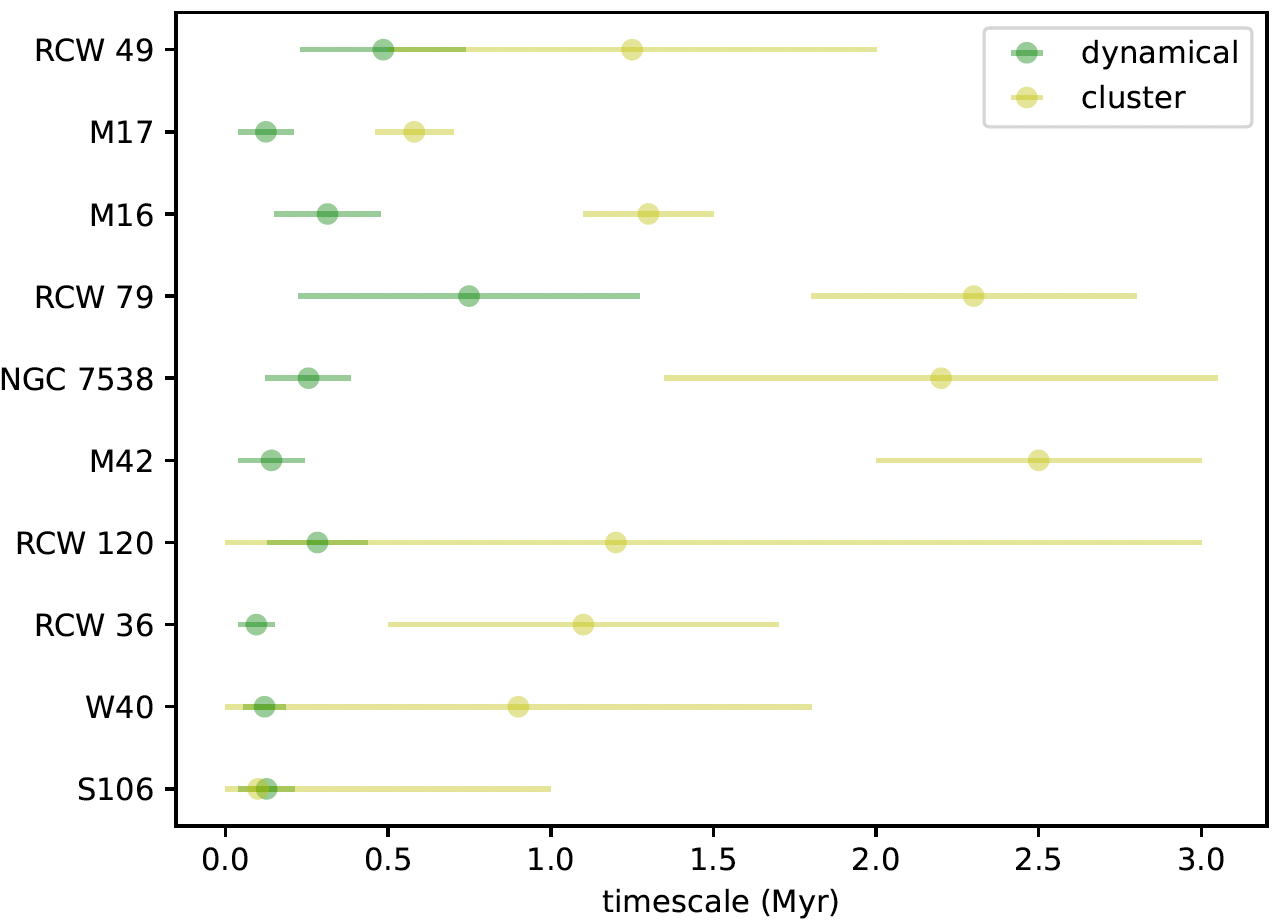}
    \caption{Stellar cluster age and uncertainty from the literature and the derived average dynamical timescale and standard deviation associated with the high-velocity gas.}
    \label{fig:dyn_cluster_ages}
\end{figure}

\begin{figure}
    \centering
    \includegraphics[width=0.95\hsize]{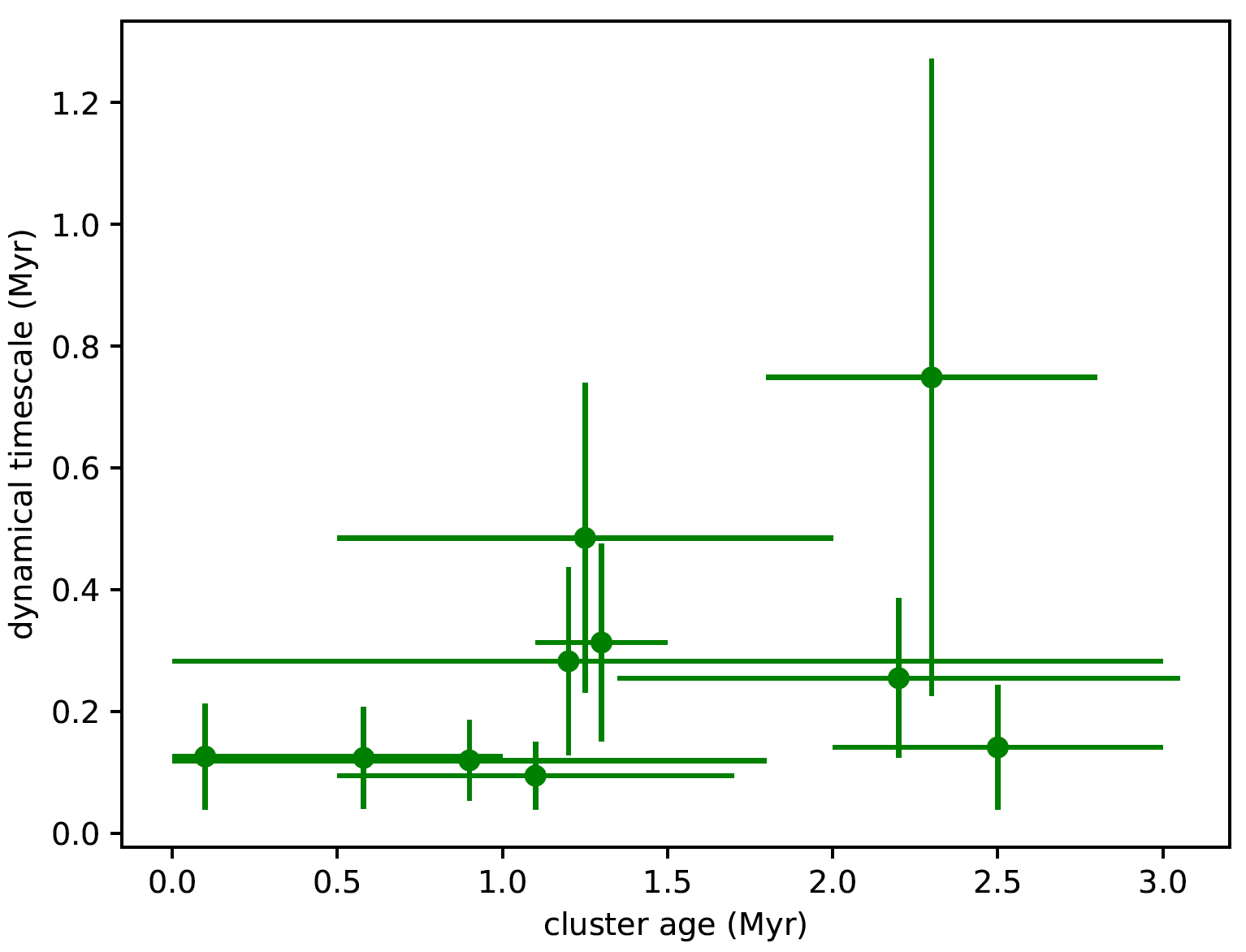}
    \caption{Derived dynamical timescales associated with the \CII\ high-velocity gas in the studied regions as a function of the stellar cluster ages from the literature.}
    \label{fig:dyn_vs_cluster}
\end{figure}

We recall that the high-velocity gas is made up of a large number of spectral bins outside the escape velocity range with respect for the spatial pixels in the map of each region.  
We thus obtained a distribution for the distances and an average velocity of the high-velocity gas for each region (Table~\ref{tab:tabObsSummary}), which we combined into a distribution for the observed t$_{dyn}$ from the \CII\ data cube. As an example, the dynamical timescale distribution for RCW79 is shown in Fig. \ref{fig:dyn_age_hists}. The distributions for the other regions are given in Fig.~\ref{fig:dyn_timescales_all}. Figure~\ref{fig:dyn_cluster_ages} summarizes the average observed t$_{dyn}$  and standard deviation for each region from these distributions alongside the corresponding cluster ages. These t$_{dyn}$, all below 0.75~Myr and typically around 0.1-0.3~Myr, are markedly shorter than the estimated cluster ages, which span approximately 0–2.5~Myr. We note that the t$_{dyn}$  are mostly in agreement with \citet{Faerber2025}, who reported values of $\sim$0.1~Myr for high-velocity expanding \CII\ shells.
With these results, we examined whether a correlation exists between the two quantities. Figure~\ref{fig:dyn_vs_cluster} presents the estimated t$_{dyn}$ as a function of cluster ages reported in the literature. No significant correlation is evident, consistent with a Pearson correlation coefficient p-value of 0.18.

\section{Analysis and discussion} \label{sec:discuss}

\subsection{Implications of the short dynamical timescales}  \label{subsec:times}
We have shown that high-velocity wings, at velocities above the cloud escape velocity, are ubiquitous in the \CII\ spectra of an unbiased sample of 10 high-mass star-forming regions with identified O stars, and that they have a dynamical origin. 
We point out that there are also regions with only identified B stars, for example M43 and NGC1977 \citep{Pabst2020} or MonR2 {(Schneider, priv. comm.), that have high-velocity wings. There are also a few regions around B stars that already have cleared most of the surrounding molecular gas, for example IC63 \citep{Bonne2023a,Caputo2023} and the Diamond Ring in Cygnus \citep{Dannhauer2025}, where the \CII\ spectra have no prominent high-velocity wings. 
Interestingly, regions that are in a very early state of evolution such as the compact \HII\ region S144 in RCW79 \citep{Keilmann2025} already show broad \CII\ wings.

The high-velocity \CII\ gas has observed t$_{dyn}$ shorter than 0.75~Myr for all regions, which is inconsistent with the stellar cluster ages in the sample that vary in age up to 2.5~Myr, as anticipated. This inconsistency can be simply demonstrated by calculating the probability that an unbiased selection of 10 targets would all have ages $<$0.75~Myr while the possible range of ages is 0 -- 2.5~Myr. The resulting probability is 5.9\,10$^{-6}$, which is smaller than the probability of a 4$\sigma$ event. From this, we learn that the observed high-velocity gas in most regions cannot be associated with a single expanding bubble or shell, proposed for M42 \citep{Pabst2019} and RCW120 \citep{Luisi2021} as in this case t$_{dyn}$ would have to be consistent, or at least directly correlated, to the stellar cluster age. 

This problem was already pointed out for RCW49 by \citet{Tiwari2021} where the authors identified an expanding shell with t$_{dyn}$ of 0.27~Myr while the existence of Wolf-Rayet stars indicate the presence of O stars for at least 2~Myr. To resolve this issue, they proposed that what we observe in RCW49 is a recent re-acceleration of the shell due to the stellar winds associated with the Wolf-Rayet stars. Unfortunately, an episodic re-acceleration scenario cannot resolve the ubiquitous presence of \CII\ high-velocity gas in all observed regions with O stars as this requires that there is high-velocity expansion or mass ejection at all stages of stellar cluster evolution. If the high-velocity gas is not present at all cluster stages, an unbiased cloud sample would have clouds where no high-velocity wings are present. The 10 sources in this study is a small sample, but we recall that in the larger sample of \citet{Faerber2025} all sources with O stars show high-velocity wings. Lastly, as the observed t$_{dyn}$ are shorter than the cluster age, the high-velocity has to be continuously replenished. 

This leaves the scenario that the bubble must break open followed by high-velocity mass ejection from the host molecular cloud through low-density or pressure holes or chimneys as described in \citet{Bonne2023c}. This also explains the distinct spatial distribution of the blue- and redshifted high-velocity gas in Fig.~\ref{fig:chanmap-rcw79} and the figures in Appendix~\ref{sec:intMaps} as the mass ejection geometry depends on the 3D geometry of the low-density holes in the turbulent host molecular cloud. The 
morphology of many of these regions already shows evidence that the bubbles have burst. RCW79 is a clear example, but also NGC 7538, M16, and RCW49. RCW120 is a special case as it has a bow structure caused by the movement of the star through the cloud. M17 is a very extreme example of an inhomogeneous cloud structure.  Note, however, that some of the associated molecular clouds, for example RCW120  \citep{Kabanovic2022}, have a flattened geometry. As a result, expanding bubbles often appear as high-contrast rings, since emission is absent along the line of sight towards the centre, making the surrounding torus appear bright.

The discrepancy of t$_{dyn}$ and cluster age is then explained by the fact that the high-velocity gas is continuously removed and pushed away from the host cloud. Good examples are RCW79 and NGC7538, where blue- and redshifted emission is detected all the way out to the edges of our observed area (that is many pc away from the cloud's centre).
Within 0.5 Myr most of the ejected gas is then no longer detected because it will be too far from the stellar cluster, preventing sufficient excitation of \CII\ emission due to lack of UV and because the local density is getting too low. Note that the bubble should burst in significantly less time than 0.5 Myr, likely on a timescale of around 0.1 Myr to explain the proposed bubble geometry in M42 and RCW120. These short timescales were demonstrated in dedicated simulations to explain the Diamond Ring \CII\ observations by sheet-like molecular cloud assembly \citep{Dannhauer2025}.

In summary, this first systematic study of high-velocity \CII\ gas thus requires a more comprehensive scenario than a simple expanding bubble to explain the observations. In Sect.~\ref{subsec:simulations} we discuss the possible processes proposed in simulations that initiate this bursting of the bubble and associated cloud erosion. 

\subsection{Previous results for the proposed scenario}  \label{subsec:previous}

We emphasize that the above scenario does not contradict the expanding bubbles proposed by \citet{Pabst2019} and \citet{Luisi2021} for M42 and RCW120, respectively, nor a number of expanding shells in other studies \citep{Tiwari2021,Beuther2022,Keilmann2025}. It instead expands the scenario, stating that the \CII\ bubbles have to break open after typically 0.1 Myr, which is close to the estimated expansion timescales for these two bubbles. In fact, \citet{Luisi2021} note potential leaking in RCW120 and for the redshifted emission they do not identify a clear expanding shell morphology (see Fig. 2 in their paper). \citet{Anderson2015} also demonstrated discontinuities in the PDR ring of RCW120 (for example their Fig.~13). 
The gas associated with this redshifted wing is thus likely already starting to stream out of the cloud through several holes that are present. In our Fig.~\ref{fig:rgb2}, we see an example in the north-east corner of the RCW120 ring where a redshifted high-velocity feature is visible. This could also explain why the integrated intensity over the high-velocity wings in RCW120 is not forming a clean circular morphology. Similarly, in \citet{Kabanovic2022} the authors argue that the observed limb brightening in RCW120 cannot be explained by a spherical geometry alone. 

The high-velocity gas in M42 on the other hand is forming a rather clean circular morphology. In addition, this region has a clear blueshifted expanding shell and very little redshifted high-velocity gas. M42 might thus be a true expanding half-bubble that will likely break open soon. There is only one puzzling point about M42. Although there is a lot of debate about the 2.5 Myr age of the stellar cluster, estimates put it far above the expansion timescale of the bubble. However, 
we note that the cluster age estimate in \citet{DaRio2010} is based on lower-mass stars. Thus, it is possible that the O stars only formed very recently in a cluster that has been forming for $\sim$2 Myr. Despite inherent difficulties in obtaining accurate ages of O stars in stellar clusters, such an argument for M42 cannot be invoked for a large number of regions because the probability is extremely low that they all would be younger than 0.6 Myr in a purely random selection process. In addition, for RCW 79 the stellar cluster age of 2.3~$\pm$~0.5 Myr was directly derived from the O star observations.

According to more detailed studies of the M42 expanding shell by \citet{Kavak2022a,Kavak2022b}, there are so-called dents in the shell, which they propose result from fossil protostellar outflows. 
Some embedded protostellar outflows were also found in RCW120 \citep{Figueira2020}, 
although not at the location where the PDR ring has broken open (see also Sect.~\ref{subsec:spatial}). Turbulence is another process that could create holes within 0.1 Myr in the molecular cloud so that gas will start streaming out, as shown in multiple turbulence simulations including stellar feedback \citep[e.g.][]{MacLow2007,Dale2012,Dale2013,Kim2018,Kim2021,Geen2021,Geen2023}. Both processes may play a role, but only further detailed studies of a large number of regions will be able to reveal the importance of both processes. 

\begin{table}[]
    \caption{Mass ejection properties of the studied regions.}
    \begin{tabular}{lcccc}
    \hline
    \hline
    {\tiny Region}  & $\dot{M}_{ej}$          & $\dot{p}_{ej}$                      & v$_{ej}$      & {\tiny wing fr.} \\
            & {\small (M$_{\odot}$ yr$^{-1}$)} & {\small (M$_{\odot}$ km s$^{-1}$ yr$^{-1}$)} & {\small (km s$^{-1}$)} &  (\%) \\
    \hline
    {\small M16 }    & 1.2\, 10$^{-2}$  & 1.2\, 10$^{-1}$ & 10.2 & 20 \\ 
    {\small M17  }   & 1.8\, 10$^{-2}$  & 3.4\, 10$^{-1}$ & 19.3 & 8 \\ 
    {\small M42  }   & 3.4\, 10$^{-3}$  & 2.9\, 10$^{-2}$ & 9.5 & 31 \\ 
    {\small NGC7538 }& 5.3\, 10$^{-3}$  & 5.8\, 10$^{-2}$ & 10.9 & 19 \\ 
    {\small RCW36}   & 3.0\, 10$^{-3}$  & 0.2\, 10$^{-2}$ & 7.0 & 21 \\ 
    {\small RCW49 }  & 1.6\, 10$^{-2}$  & 3.0\, 10$^{-1}$ & 19.5 & 14 \\ 
    {\small RCW79 }  & 1.0\, 10$^{-2}$  & 9.8\, 10$^{-2}$ & 9.6 & 28 \\ 
    {\small RCW120 } & 4.6\, 10$^{-3}$  & 5.1\, 10$^{-2}$ & 11.1 & 36 \\ 
    {\small W40 }    & 1.5\, 10$^{-3}$  & 1.2\, 10$^{-2}$ & 7.9 & 24 \\ 
    {\small S106 }   & 1.8\, 10$^{-3}$    &  1.1\, 10$^{-2}$ & 6.5  & 44 \\
    \hline
    \end{tabular}
    \tablefoot{The average mass and momentum ejection rates, assuming n$_{H}$ = 5\, 10$^{3}$ cm$^{-3}$, the average ejection velocity and the fraction of the wings to the average spectrum. Error estimates for these values are difficult, but based on the discussion of the assumptions that go into the calculations a factor of 2 should be considered for the uncertainties on the mass and momentum ejection rates.
    }
    \label{tab:massEjTable}
\end{table}

\begin{figure}
    \centering
    \includegraphics[width=0.85\hsize]{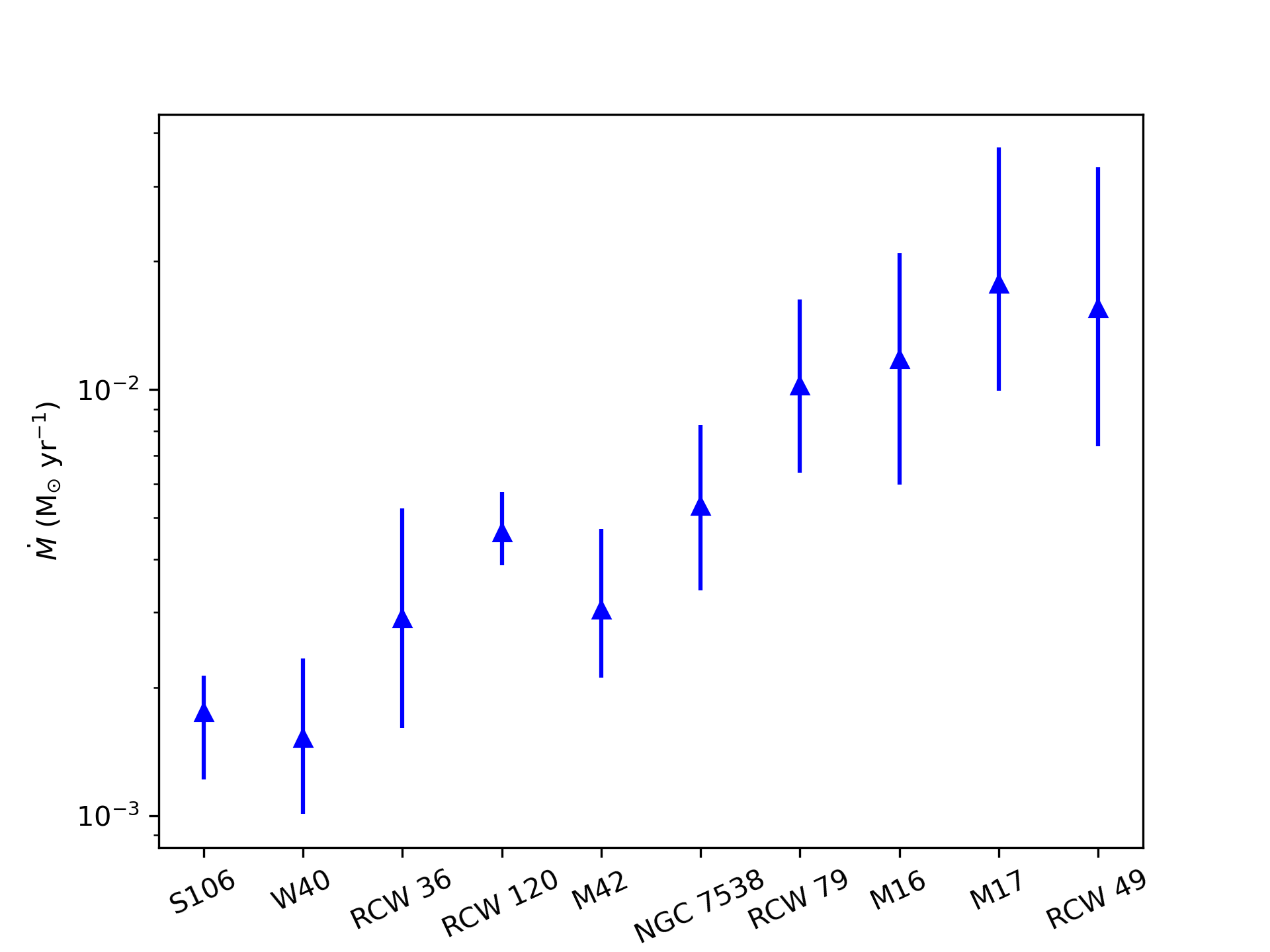}
    \includegraphics[width=0.85\hsize]{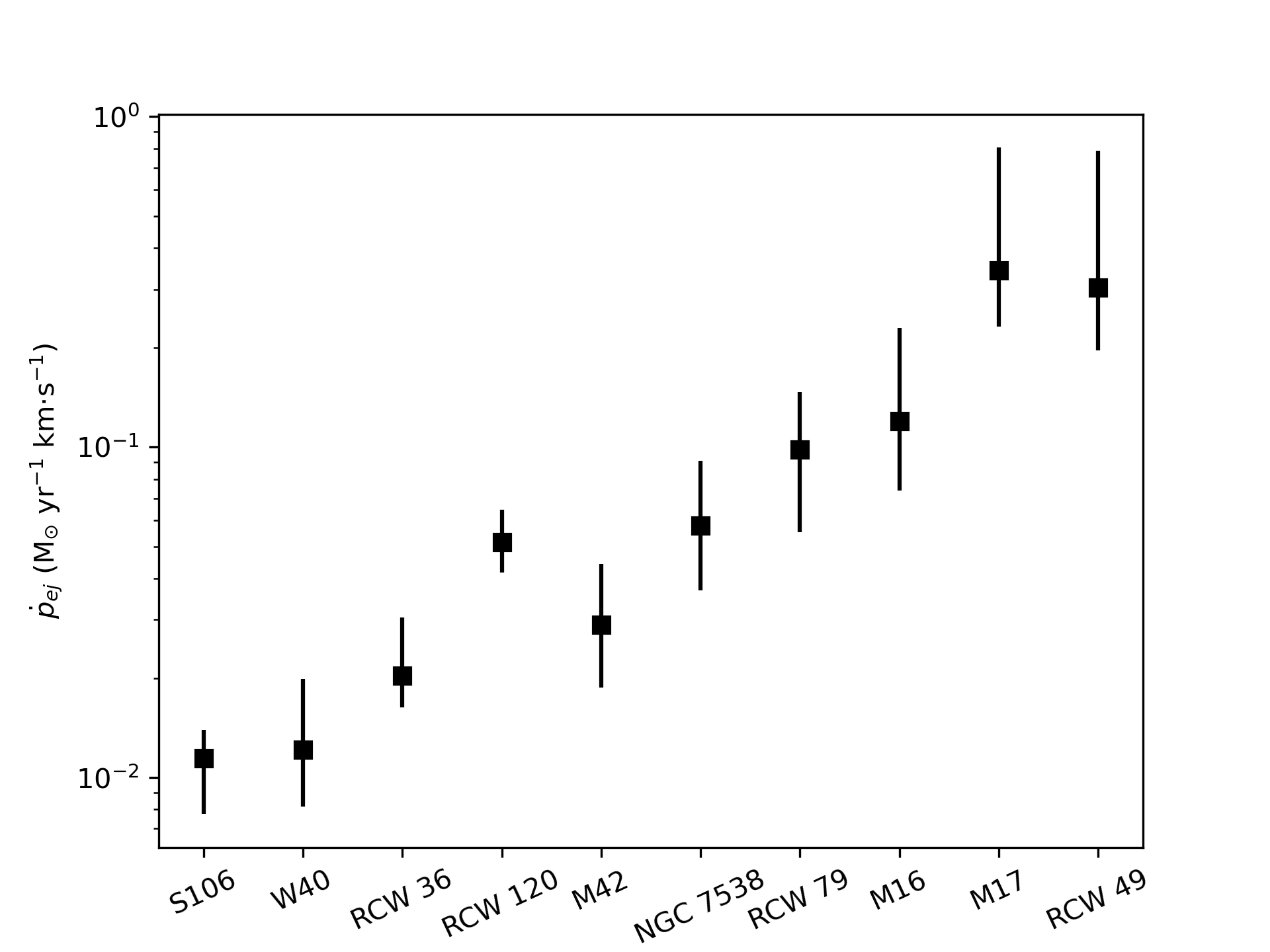}
    \includegraphics[width=0.85\hsize]{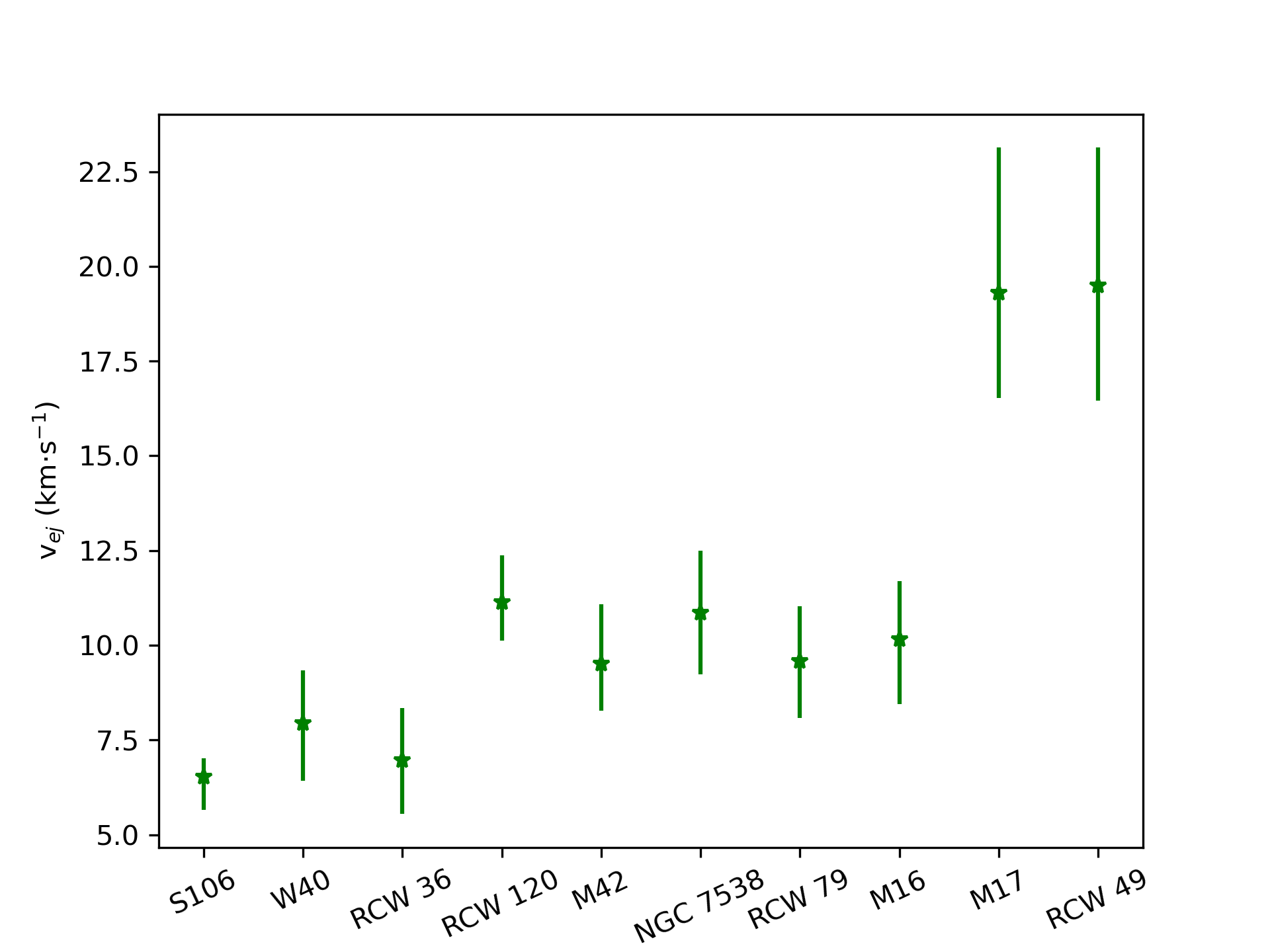}
    \caption{Mass ejection properties. The three panels show the estimated mass ejection rate (\textit{top}), momentum ejection rate 
    (\textit{middle}), and average mass ejection velocity (\textit{bottom}) for the studied regions. The number and luminosity of the exciting O stars increase from left to right.  
    }
    \label{fig:mej_rates_fig}
\end{figure}

\subsection{Mass ejection rates from molecular clouds} \label{subsec:mass-ejection}
If we presume that the \CII\ high-velocity gas results from mass ejection from the host molecular cloud, it raises 
the question how effective stellar feedback is in removing mass from the host cloud. 
To quantify the mass ejection rates for our clouds, we built on the work presented in \citet{Bonne2023c}.
First, the C$^{+}$ column density was calculated from the integrated intensity of the high-velocity wings, assuming that the wings are optically thin, using the equations from \citet{Goldsmith2012}:
{\small
\begin{equation}
    \Delta T_{A} = 3.43\,10^{-16} \times\left[ 1 + 0.5\, {\rm e}^{91.25/{\rm T_{kin}}} \left( 1 + \frac{2.4\times10^{-6}}{{\rm C_{ul}}} \right) \right]^{-1} \frac{{\rm N(C^{+})}}{\delta {\rm v}}
,\end{equation}
}with T$_{kin}$ the kinetic temperature, N(C$^{+}$) the C$^{+}$ column density, $\delta$v the spectral bin width, and C$_{ul}$ the collisional de-excitation rate given by 
$    C_{ul} = n \times R_{ul} $,
where n is the density and R$_{ul}$ is the de-excitation rate coefficient for atomic hydrogen given by
\begin{equation}
    R_{ul} = 7.6 \, 10^{-10}\,{\rm cm^{3}}\,{\rm s^{-1}}\,\left(\frac{T_{kin}}{100\,{\rm K}}\right)^{0.14}
.\end{equation}
For the kinetic temperature (T$_{kin}$), we assumed a value of 100 K and a hydrogen density n value of 5\,10$^{3}$ cm$^{-3}$. 
These are justified assumptions, based on theory \citep{Goldsmith2012} and the experience of many studies of \CII\ emission in the last years, covering a large range of physical properties for the molecular cloud and PDR \citep[e.g.][]{Schneider2018,Luisi2021,Pabst2022,Tiwari2022,Bonne2022b,Bonne2023c,Schneider2025}.
For the collisional partners (H, H$_{2}$, and e$^{-}$), we considered that atomic hydrogen as \CII\ mostly originates in the neutral ISM \citep{Tielens1985,Pineda2013,Schneider2023} and that the wings have no counterpart in CO (see Sect.~\ref{subsec:spatial}). 
The optically thin assumption appears to generally hold in the high-velocity wings \citep{Kabanovic2022}. For the kinetic temperature, several studies of different FEEDBACK regions with PDR models have pointed to temperatures around 100 K, varying between 50 and 500 K \citep[see e.g.][]{Tiwari2022,Bonne2022b,Bonne2023c}. \citet{Bonne2022b} showed that these temperature variations maintain column density uncertainties within a factor of 2. The density is more difficult to constrain. Taking n=n$_{H}$ = 5\,10$^{3}$ cm$^{-3}$ is  justified assuming thermal equilibrium with a nearby ionized gas phase with an electron density of the order of n$_{e}$ $\approx$ 10$^{2}$ cm$^{-2}$. In addition, densities from other \CII\ studies (see above) point towards a value of typically a few 1000 cm$^{-3}$. Obviously, the assumed density has an impact on the derived mass ejection rate. Reducing the density to n$_{H}$ = 5\,10$^{2}$ cm$^{-3}$ increases the mass by more than a factor of 3. On the other hand, increasing the density to n$_{H}$ = 5\,10$^{4}$ (5\,10$^{5}$) cm$^{-3}$ only reduces the mass by $\sim$20\% ($\sim$25\%). However, at these high densities, atomic hydrogen will be mainly in molecular form.

The C$^{+}$ derived column density map was then converted to a hydrogen column density map using the [C$^{+}$]/[H] = 1.6\,10$^{-4}$ abundance ratio reported in \citet{Sofia2004} from which the mass was calculated, considering the 1.36 correction factor due to heavier elements. The resulting mass ejection rate $\dot{M}_{ej}$ is then calculated from the total mass M$_{wings}$ as a sum over the high-velocity wings and the average t$_{dyn}$ with $\dot{M}_{ej}$ = M$_{wings}$/t$_{dyn}$. The momentum rate ($\dot{p}_{ej}$) is determined by multiplying 
$\dot{M}_{ej}$ by the average ejection velocity (v$_{ej}$), which is calculated from the weighted average of the spectral wings multiplied by $\sqrt{3}$ to correct for average projection effects.

Table \ref{tab:massEjTable} presents the mass ejection rates, the average ejection velocity, and momentum rate.  Because of the uncertainties on the density in the calculations, the presented mass and momentum ejection rates could be considerably higher if lower-density gas is significantly contributing to the observed emission.\footnote{The mass ejection rate of RCW79 is at least a factor of 2 smaller than the value derived in \citet{Bonne2023c} because we used a different mass and dynamic timescale estimate.}
Inspecting the mass ejection rates we observe that they have range of 10$^{-3}$ - 10$^{-2}$ ${M}_{\odot}$ yr$^{-1}$ and vary from region to region. First of all, it is interesting to note that these mass ejection rates have a magnitude similar to the estimated mass inflow rates in gravitationally collapsing high-mass clouds and clumps \citep[e.g.][]{Schneider2010,Peretto2013,Peretto2014,He2015}. This suggests that cloud erosion by stellar feedback is capable of balancing proposed gravitational inflow in the clumps where high-mass stars form. Provided that mass ejection is continuous, this shows that stellar feedback has the ability to play a significant role in regulating the star formation rate in gravitationally collapsing clouds and clumps. To  evaluate the impact of stellar feedback, it is important to study these quantities as a function of the stellar cluster properties. For a more holistic view of stellar feedback this should also include the mass and column density of the molecular cloud. However, this is out of the scope of this paper as this requires significant further analysis of the cloud and cluster sample. In addition, the current sample size is still limited for a proper study of scaling laws. 

The stellar cluster luminosities are not readily available in the literature, but for all clusters there is a fairly good characterization of the dominant O stars spectral types (Table~\ref{tab:tabObsSummary}), ranging from regions such as W40 and RCW36 with one or two O9V type stars up to RCW49, which is one of the brightest \HII\ regions in the Milky Way \citep{DeBuizer2022} and contains Wolf-Rayet stars. 
This classification is now used for a qualitative evaluation of the correlation between the mass ejection rate, momentum, and velocity and the cluster type; this is shown in Fig.~\ref{fig:mej_rates_fig}, roughly ordered as a function of the number of O-stars of the cluster. We emphasize that this not an absolute cluster luminosity ranking and we refrain from using quantities like the total 70 $\mu$m flux or the Lyman continuum flux because they only assess the radiative impact and not the one from stellar winds. 
In future studies, we will explore in more detail and with a larger statistics of sources, which quantitative measures could be employed. The figure illustrates that all properties,  mass ejection rate, momentum rate and ejection velocity, increase with the estimated stellar content. This finding is anticipated because larger stellar clusters with brighter O-stars should inject more energy in the surrounding medium. Interestingly, M16 deviates slightly from the general trend, partly caused by the lower ejection velocity, compared to M17 and RCW49. 

\subsection{The dispersal times of molecular clouds} \label{subsec:dispersal}

With the available cloud mass and mass ejection rates, we estimated the dispersal timescale of the cloud and the lifetime of these molecular clouds once the first O stars have formed.\footnote{Note that this approach is different from those of \citet{Bertoldi1990}, \citet{Lefloch1994}, \citet{Johnstone1998}, and \citet{Whitworth2004}, who provide analytic models for the evaporation of dense gas clouds by sources of UV radiation.} 
We assumed there is no erosion until cluster formation and that the current mass ejection rate in each cloud is constant and  representative of the average mass ejection rate over the lifetime of the O star or molecular cloud. The dispersal timescale of the cloud is defined as the mass of the cloud (see Table~\ref{tab:tabObsSummary}) divided by the observed mass ejection rate determined with our \CII\ observations (Table \ref{tab:massEjTable}). Adding the estimated cluster age then results in the estimated cloud lifetime once the first O stars form. From Fig. \ref{fig:lifetime}, we observe that the erosion timescale and cloud lifetime are typically of the order of 3-10 Myr, which is  consistent with multiple indirect observational results \citep[e.g.][]{Leisawitz1988, Leisawitz1989, Hannon2019, KimCh2021, KimCh2025, Chevance2022}. Figure \ref{fig:lifetime} does show significantly longer cloud erosion timescales up to 25-30 Myr for NGC 7538 and M17. It is important to note that using the CO-based cloud mass estimates for M17 \citep{Elmegreen1979, Povich2009} brings cloud erosion timescales back into a range of 8 - 16 Myr, which is close to the values observed for most other clouds. Similarly, it might be that our \textit{Herschel}-based mass estimate for the NGC7538 molecular cloud overestimates its mass. From Table \ref{tab:tabObsSummary}, it becomes obvious that M17 and NGC7538 have the largest effective cloud radius. Additional clouds in the LOS might thus contribute to the mass. However, if the mass estimates in this paper for M17 and NGC7538 prove to be correct it suggests there might be large molecular clouds with lifetimes up to 30 Myr. Of course, additional high-mass star formation in the densest regions of these giant molecular clouds might eventually lead to increased mass ejection and a reduced molecular cloud lifetime.

Lastly, we note that all clouds, including M17 and NGC7538, appear capable to erode all the molecular gas within a radius of 10 pc in less than 7 Myr. Even though for some clouds it could take longer to erode the full molecular cloud, all clouds are capable to erode all the gas in the high-mass star-forming clumps that are proposed to be subject to gravitational collapse \citep{Peretto2023}. Stellar feedback is thus an effective mechanism to maintain a relatively low star formation rate, even when gravitationally collapse initiates high-mass star formation. Interestingly, this timescale is very similar to the timescale of a first supernova to happen (i.e. typically 3 to 10 Myr for O stars).

\begin{figure}
    \centering
    \includegraphics[width=0.95\linewidth]{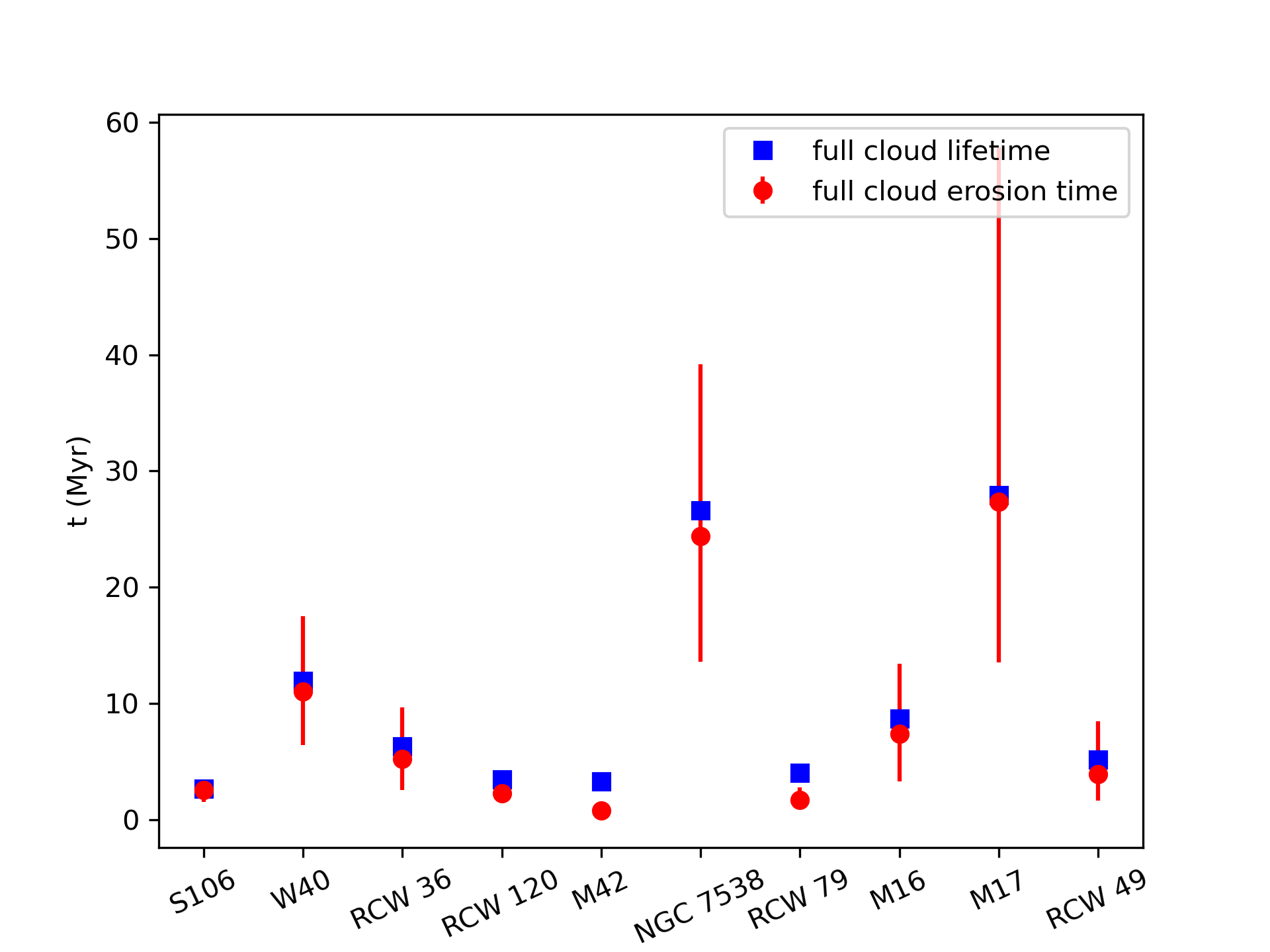}
    \caption{Cloud erosion time and the estimated full cloud lifetime (i.e. erosion time plus cluster age). The cloud erosion time is the total cloud mass divided by the observed mass ejection rate.}
    \label{fig:lifetime}
\end{figure}

In the next section, we discuss these results in the context of predictions by recent turbulence simulations including stellar radiation feedback.

\subsection{Comparison with stellar feedback simulations} \label{subsec:simulations}
As shown by \citet{Bonne2023c}, the mass ejection rates inferred for RCW79 are in good agreement with predictions from simulations designed to reproduce an RCW79-type \HII\ region \citep{Walch2012}. 
These authors performed smoothed particle hydrodynamics simulations considering only radiation, but different fractal dimensions of the associated molecular cloud that has a fixed mass of 10$^4$ M$_\odot$. However, here we are confronted with a cloud sample covering a large range of masses and radii and one-to-one comparisons are difficult.  

In general, stellar feedback models can be divided into three categories, those that consider only radiation \citep[e.g.][]{Dale2005, Krumholz2006, Krumholz2009,Walch2012,Geen2015}, only stellar wind \citep[e.g.][]{Dale2008, Dale2013b,Wareing2017a,Wareing2017b,Wareing2018}, and both \citep[e.g.][]{Dale2017,Geen2021,Geen2023,Lancaster2025b}. Some models include the role of self-gravity or magnetic fields and supernova explosions. 
Radiative feedback involves direct radiation pressure and dust-processed radiation pressure, which are generally not considered to play a dominant role in the dynamics of \HII\ regions \citep[e.g.][]{Matzner2002,Lopez2014}, as well as the more significant pressure of warm ionized gas. The predictions from models can be quite different. For example, the initial cloud structure has a major influence on the effectiveness on stellar feedback \citep[e.g.][]{Walch2012,Lancaster2021a}. \citet{Krumholz2006} consider the properties of the \HII\ region mostly insensitive to local inhomogeneities. 

It is challenging to model the process of gas removal itself and unfortunately, simulations do not always provide predictions that can be compared with observationally derived quantities in this work. In addition, a proper comparison with numerical simulations should include radiative transfer to constrain multiple observational biases that go into the estimates presented above. 
Some qualitative comparisons are, however, possible. 

Numerical simulations find that the formation of low-density channels through which the ejected gas can escape reduces the pressure in the thermal \HII\ region. However, this does not necessarily prevent rapid cloud erosion in these simulations as this does not lead to a reduction of the mass ejection rate \citep{Walch2012,Dannhauer2025}. Mass can thus be continuously ejected at the interface of the molecular cloud (e.g. through the rocket effect) while lower-density channels form and evolve. 
For radiation-only simulations, \citet{Dale2013} for example find that the more massive and denser the clouds are, they remain largely unaffected by ionization, while for the lower-mass and lower-density clouds a substantial fraction of their gas reserves is expelled. They emphasize that the reaction of the clouds to photoionization depends  strongly on their escape velocities.  We can more quantitatively compare the observed mass ejection velocities with the results from the grid of simulations presented in \citet{Kim2018}. They found mean neutral mass ejection velocities of $\sim$7-15~km~s$^{-1}$ and mean ionized mass ejection velocities of $\sim$20-35~km~s$^{-1}$. The observed values with \CII\ are spread over this range of values. This could fit the idea that the \CII\ high-velocity wings trace both neutral and ionized mass ejection from the regions. However, unlike the observations, the mean mass ejection velocities in the simulations by \citet{Kim2018} do not show a trend with the formation of more massive clusters. This could indicate a discrepancy with these simulations, but we should also consider a bias in our definitions. In \citet{Kim2018} the mean mass ejection velocity is calculated based on all the gas above the escape velocity from the cloud. In our work we have a similar definition, but in some regions (RCW79, M16, M17, and RCW49) we adjusted this definition and used a higher escape velocity threshold to exclude contributions from additional CO velocity components. As these are the most luminous star clusters, this could create an artificial trend in our observations. Removing these four regions from the sample reduces the significance of the mass ejection velocity trend, but there remains a trend from about 7~km~s$^{-1}$ to 11~km~s$^{-1}$ from regions with a single O9 star to regions with multiple O stars.

Concerning wind-only or radiation- and wind combined simulations, the results of the models differ. While \citet{Dale2017} state that photoevaporation is the most important process in the disruption of molecular clouds, \citet{Wareing2017a, Wareing2017b} emphasize the importance of stellar winds and their wind-only models efficiently manage to qualitatively reproduce our observations of  cloud structure and high-velocity emission. In particular, their sheet models of molecular cloud structure \citep{Wareing2017a, Wareing2018} that also include magnetic fields closely match our observations, i.e. most of our observed molecular clouds are not spherical but rather flat or sheet-like (see also \citealt{Beaumont2010, Dannhauer2025}) and that high-velocity gas can escape through evacuated cavities and elongated stellar wind tunnels. 
\citet{Geen2023} promote a scenario in which stellar wind bubbles mix with the photoionized gas when the \HII\ region breaks out of the cloud as a champagne flow, and dub this process ‘hot champagne’. 

\section{Summary} \label{sec:conclusions}
We performed a systematic study of \CII\ 158 $\mu$m emission around ten high-mass star-forming regions that are excited by at least one O star using spectrally resolved  data from SOFIA (FEEDBACK and the C+SQUAD Orion legacy surveys and archive data). 
We find that the \CII\ spectra in all these star-forming regions have high-velocity wings that significantly (8\% - 44\%) contribute to the total \CII\ luminosity of the region. These high-velocity wings have  complex spatial distributions, and there is evidence that they stream out of the cloud in areas where the morphology and location of the blue- and redshifted high-velocity gas can be entirely different. There is also regular blue-red asymmetry when it comes to the integrated intensity of these high-velocity wings. The high-velocity wings have a typical observed dynamical timescale of 0.1-0.3 Myr, with maximal values reaching 0.75 Myr. 
This creates an important tension with the determined ages of O star clusters, 
which are all below 3 Myr and have a strong variation in their age. This, combined with the complex spatial distribution, points to a scenario where expanding bubbles break open in $\sim$0.1 Myr to a few times 0.1 Myr, after which the high-velocity gas streams out of the molecular cloud through low-density holes. Quantifying the associated mass ejection rates provides lower limits starting around 10$^{-3}$~M$_{\odot}$~yr$^{-1}$ for RCW36 and W40; this limit is  $>$10$^{-2}$~M$_{\odot}$~yr$^{-1}$ for M17 and RCW 49. These mass ejection rates result in a cloud lifetime for most regions of 3-10 Myr after the first O stars form, consistent with multiple other studies. These observed mass ejection rates are also remarkably similar to observed mass inflow rates at the early stages of high-mass star-forming regions. This demonstrates that stellar feedback can balance this mass inflow and thus provides compelling evidence that stellar feedback may play a significant role in maintaining the low observed star formation rates, even when high-mass star formation is driven by cloud-scale gravitational collapse.

\begin{acknowledgements}
We thank the anonymous referee for constructive and insightful comments that improved the clarity of this paper. 
This study was based on observations made with the NASA/DLR  Stratospheric Observatory for Infrared Astronomy (SOFIA). SOFIA is
jointly operated by the Universities Space Research Association Inc. (USRA), under NASA contract NNA17BF53C and the Deutsches SOFIA
Institut (DSI), under DLR contract 50 OK 0901 to the University of Stuttgart. upGREAT is a development by the MPIfR and the KOSMA/University Cologne, in cooperation with the DLR Institut f\"ur Optische Sensorsysteme.
Financial support for FEEDBACK at the University of Maryland was provided by NASA through award SOF070077 issued by USRA. The FEEDBACK project was supported by the BMWI via DLR, project number 50OR2217. 
S.D. acknowledges support from the International Max Planck Research School (IMPRS) for Astronomy and Astrophysics at the Universities of Bonn and Cologne. S.K. acknowledges support from the BMWI via DLR, project number 50OR2311, and funding from the DFG project number 558818801. This work was supported by the CRC 1601 (sub-projects B1, B2) funded by the DFG – 500700252.
\end{acknowledgements}

\bibliography{aa59930-26}

\begin{appendix}
\onecolumn
\section{Average spectra of the FEEDBACK regions}\label{sec:avSpec}

In Fig. \ref{fig:avSpecs}, the average \CII\ and CO spectra from the FEEDBACK sources (M16, M17, NGC 7538, RCW36, RCW49, RCW120, and W40), M42, and S106 are presented. For the FEEDBACK sources, the $^{12}$CO and $^{13}$CO(3-2) lines are shown, except for  NGC7538 and M42, where the $^{12}$CO and $^{13}$CO(2-1) lines are displayed, and S106, where the $^{13}$CO(2-1) line is shown. 

\begin{figure*}[h]
    \centering
    \includegraphics[width=0.32\hsize]{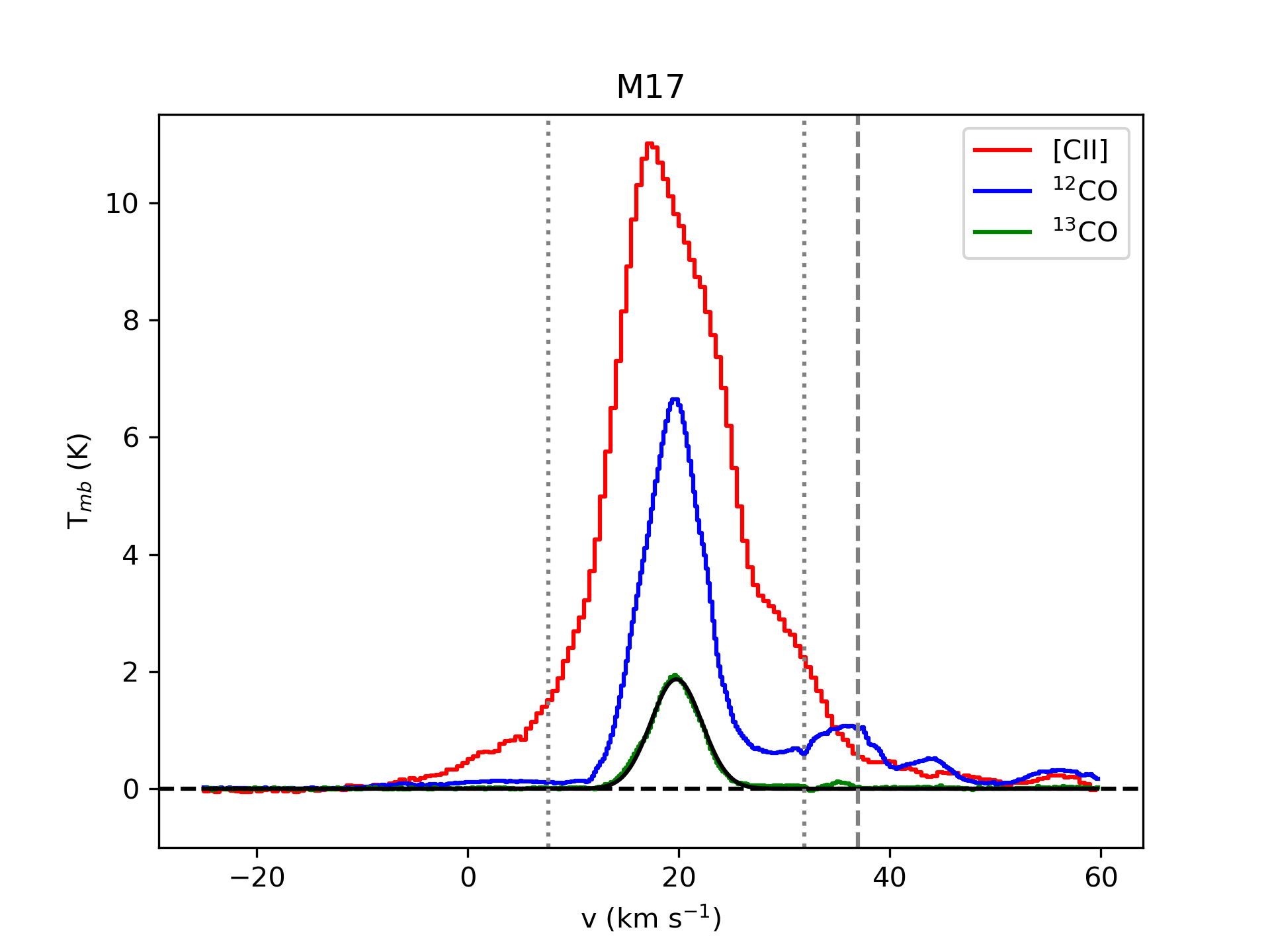}
    \includegraphics[width=0.32\hsize]{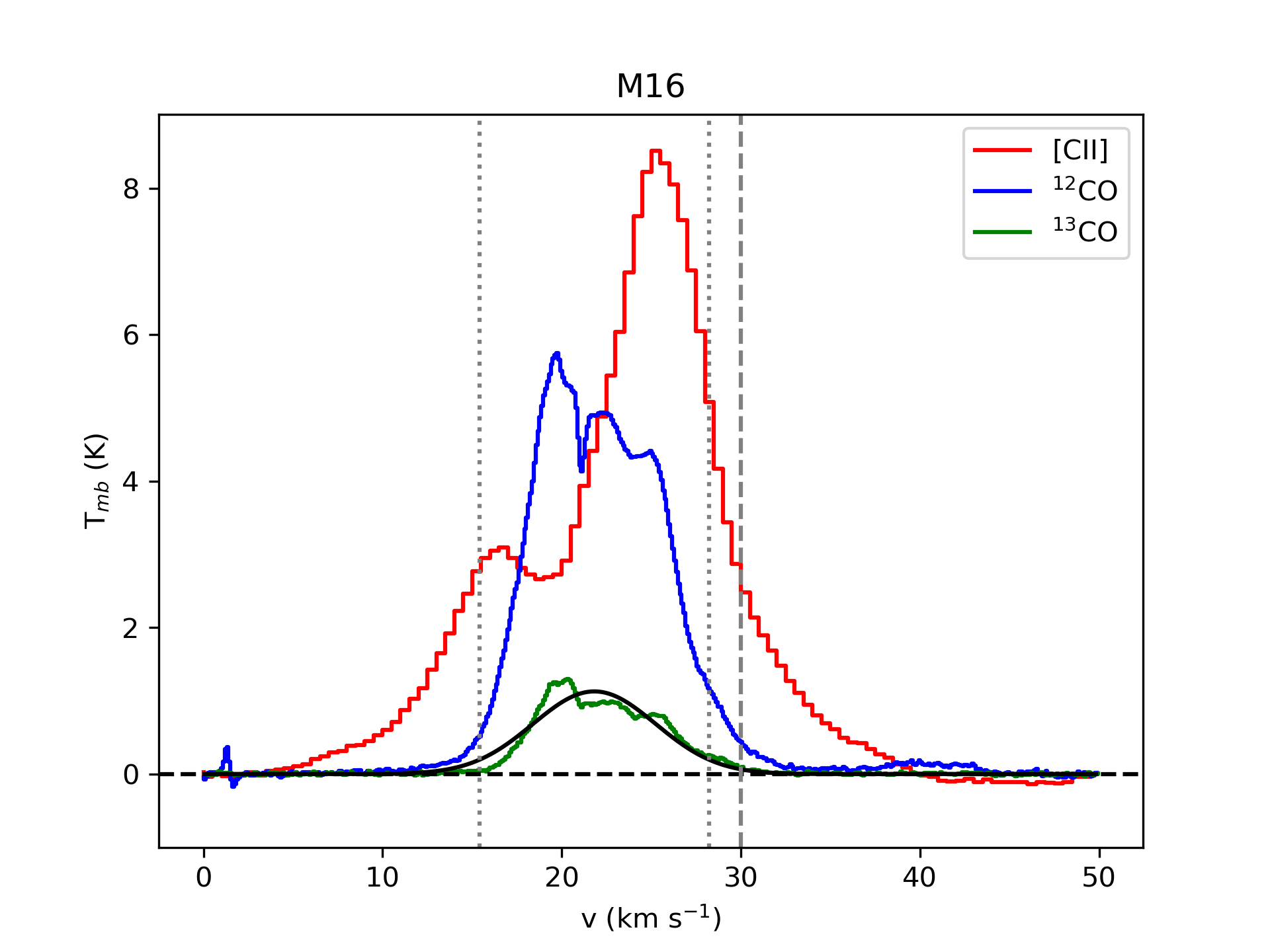}
    \includegraphics[width=0.32\hsize]{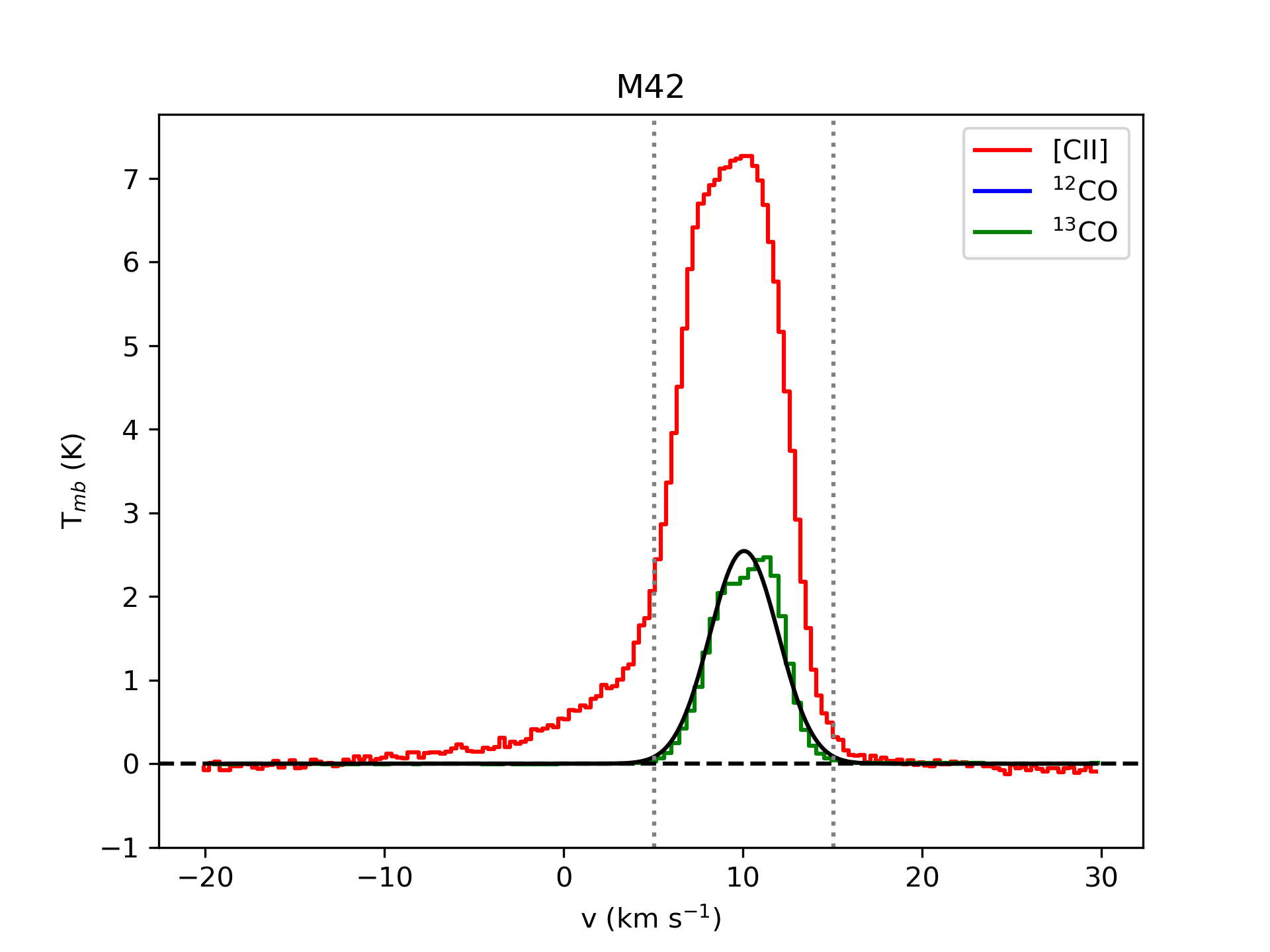}
    \includegraphics[width=0.32\hsize]{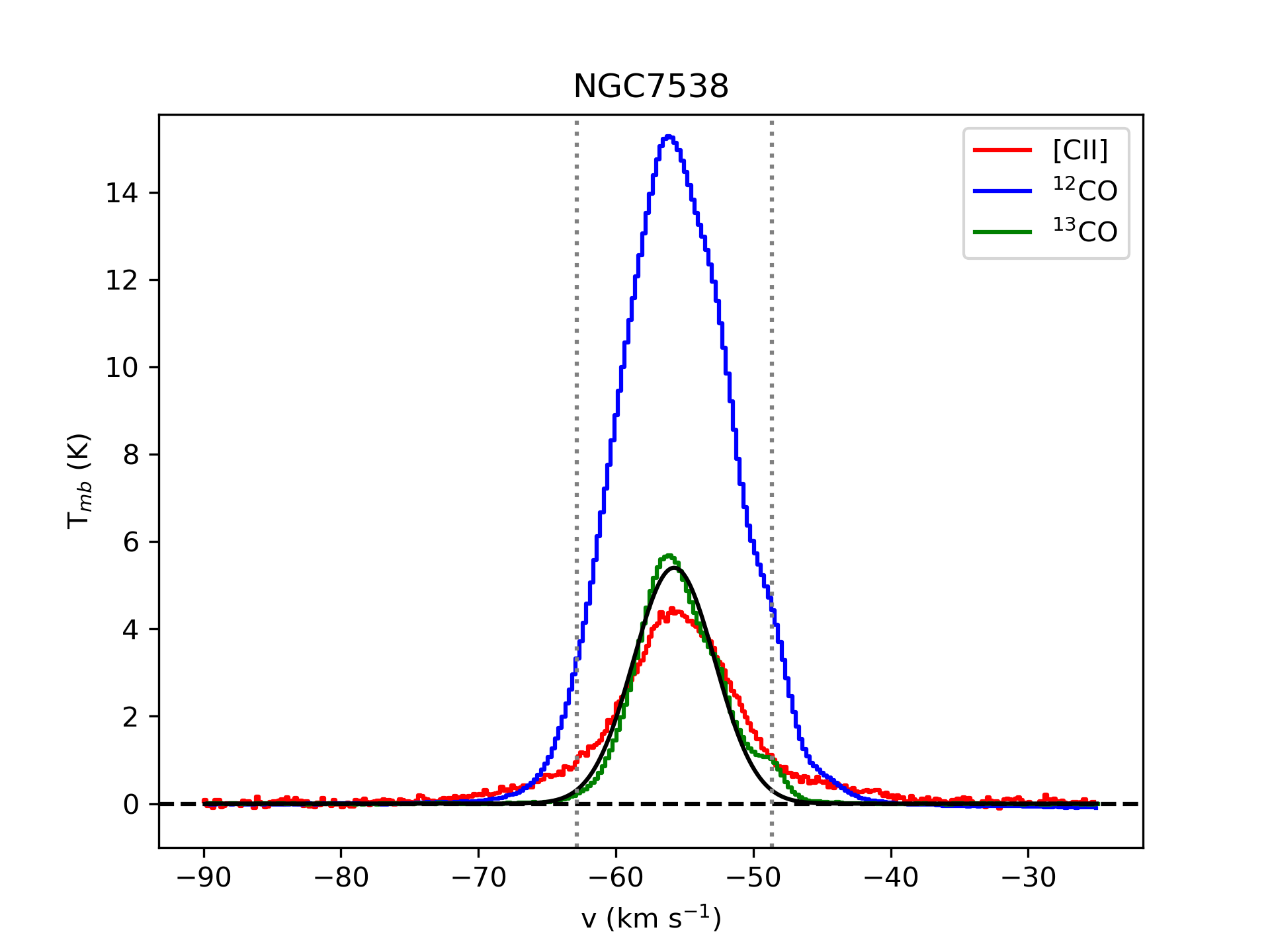}
    \includegraphics[width=0.32\hsize]{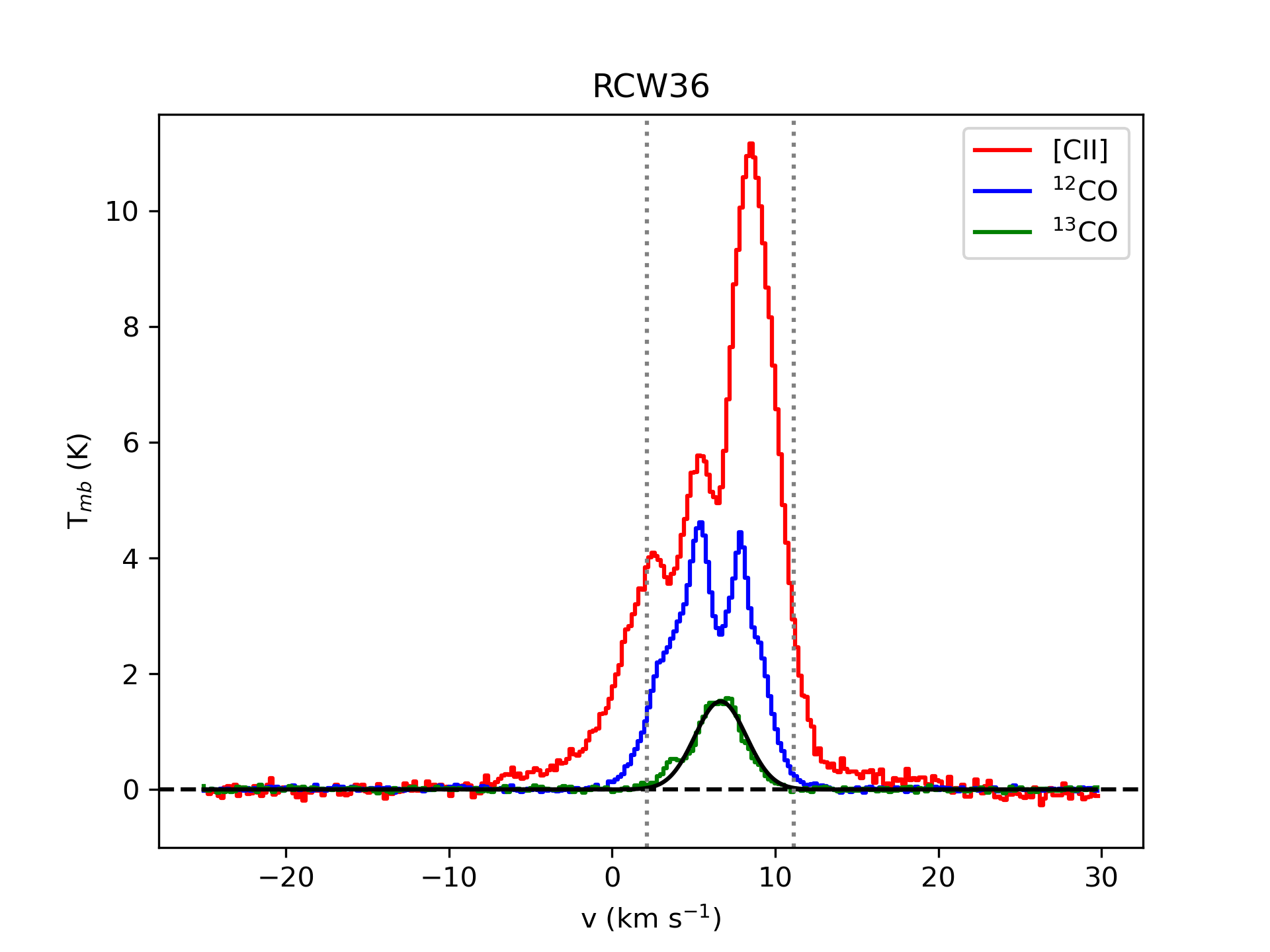}
    \includegraphics[width=0.32\hsize]{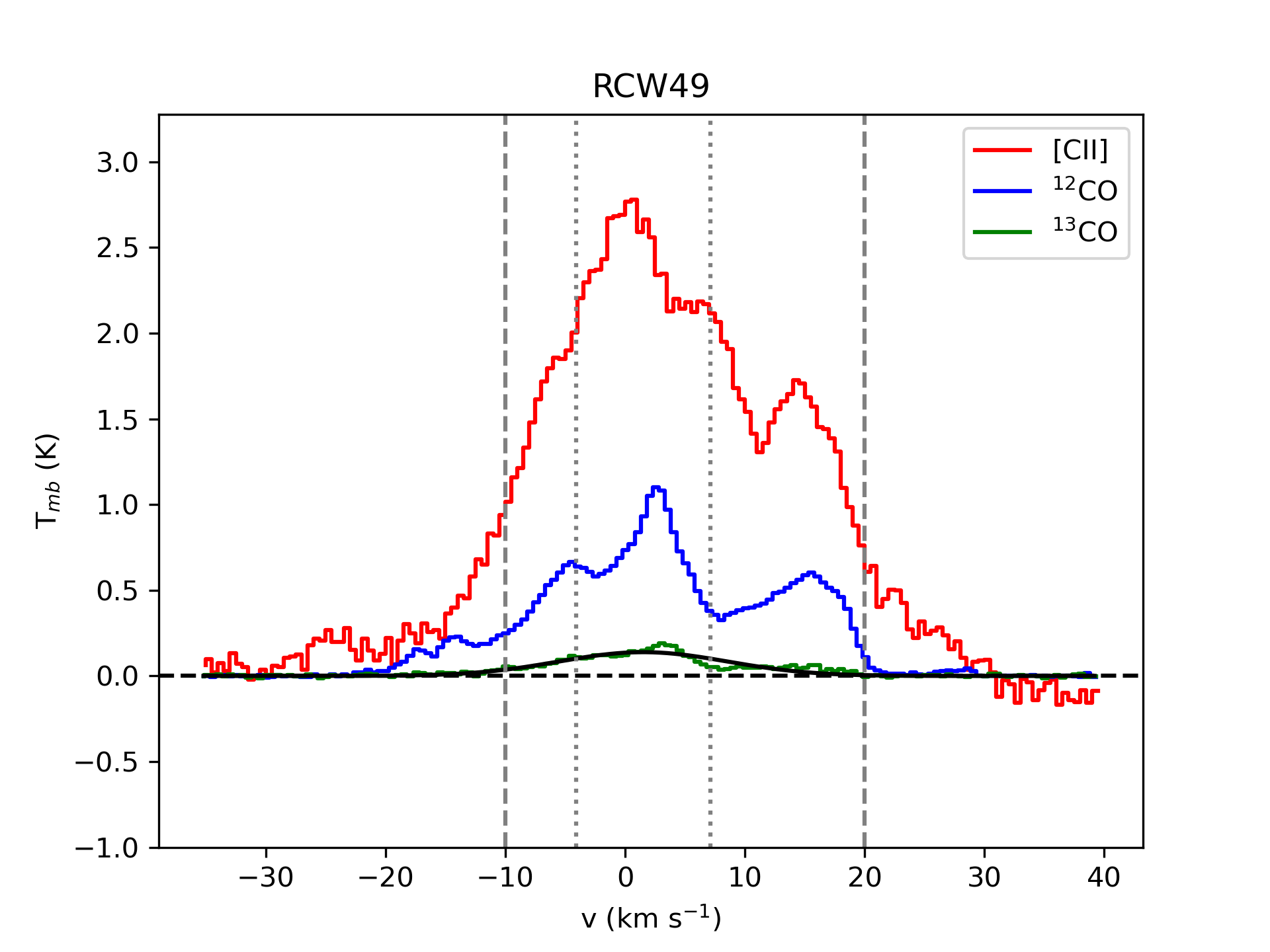}
    \includegraphics[width=0.32\hsize]{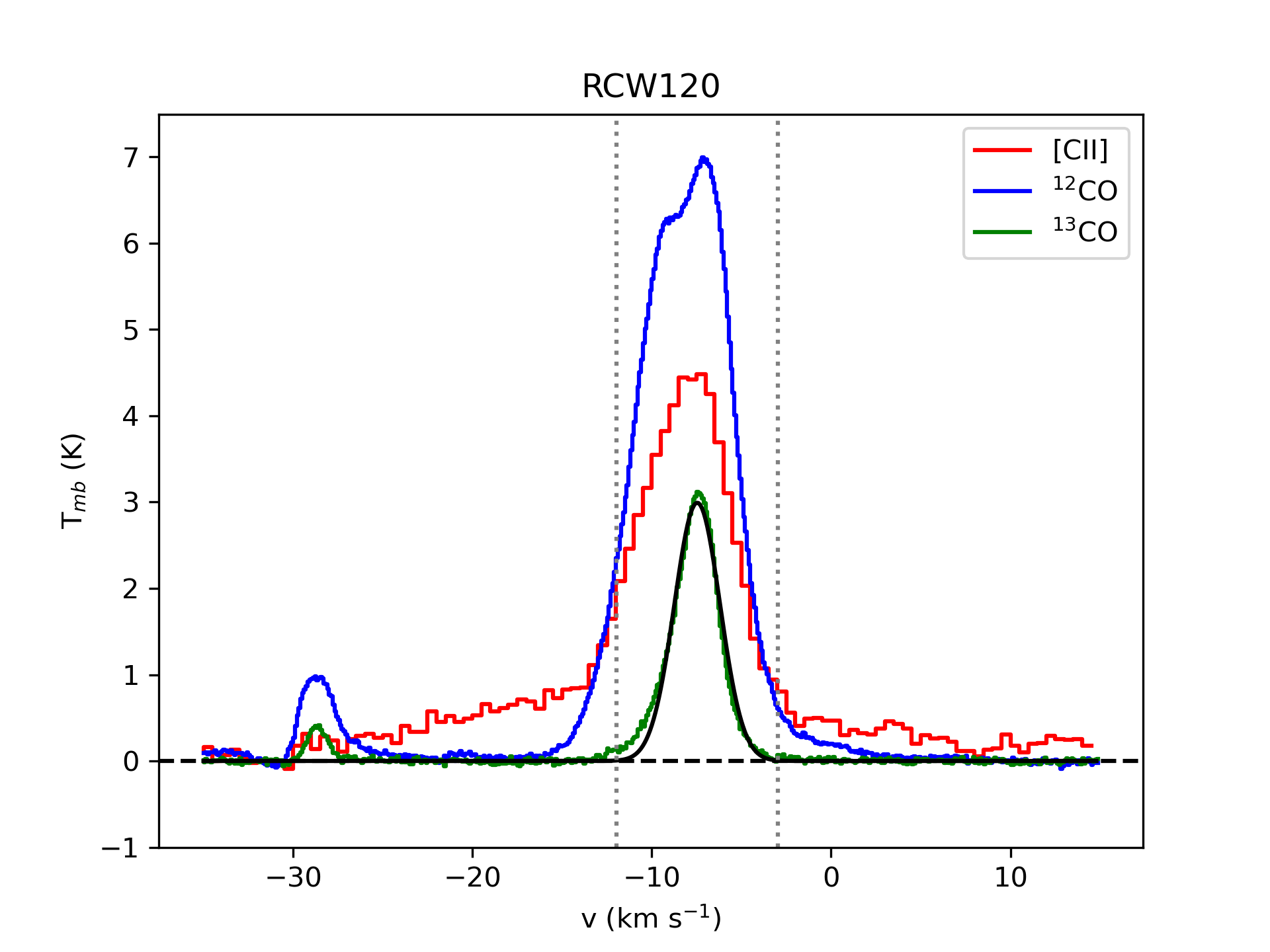}
    \includegraphics[width=0.32\hsize]{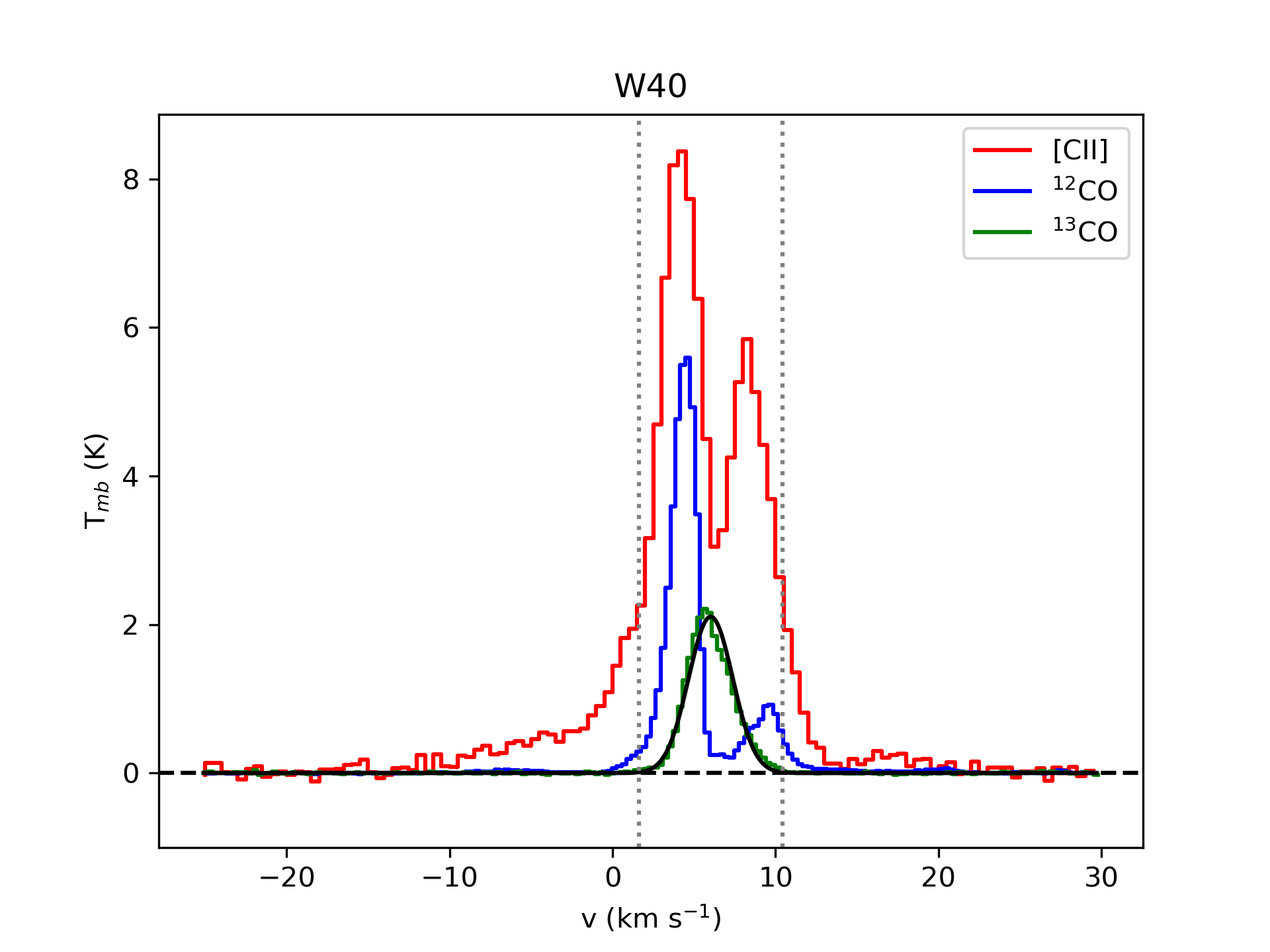}
    \includegraphics[width=0.32\hsize]{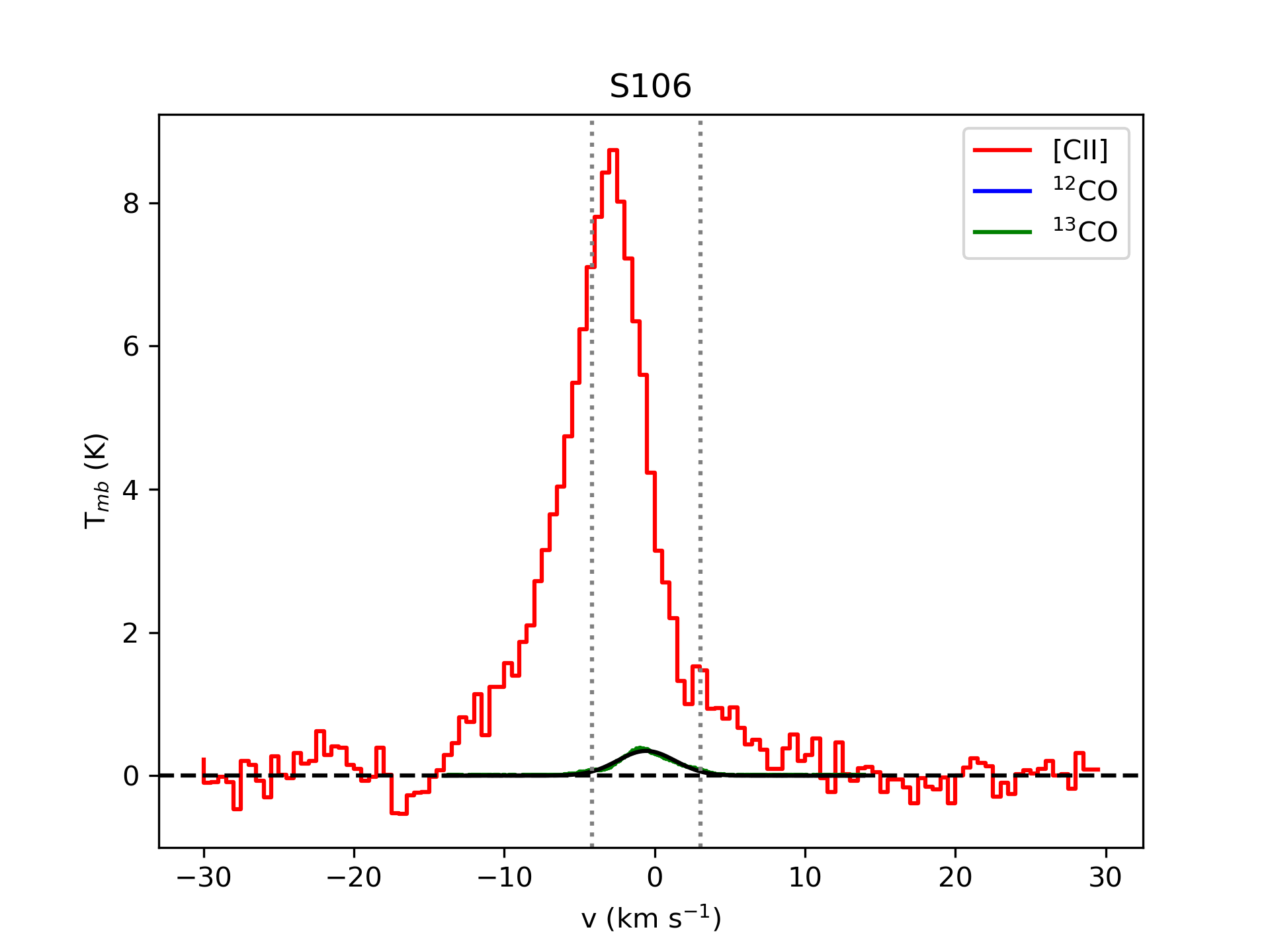}
    \caption{Average \CII\ and CO spectra for the regions that are not presented in Fig. \ref{fig:avSpecM17-RCW79}. 
    The vertical dotted lines mark the velocity intervals associated with the high-velocity wings, while the vertical dashed lines indicate the revised threshold for the onset of high-velocity emission, chosen to exclude contamination from fore- or background sources. The Gaussian fit to the $^{13}$CO line is in black.}
    \label{fig:avSpecs}
\end{figure*}
\newpage

\section{Integrated intensity RGB maps of the FEEDBACK regions}\label{sec:intMaps}
In Figs.~\ref{fig:rgb1} to \ref{fig:rgb5}, the integrated \CII\ intensity (moment-0) RGB maps are shown together with overlays of CO emission. To improve the quality of the moment-0 maps of the individual colour channel, a moment-masking was applied, following the general approach of \citet{Adler1992} and described in more detail in \citet{Dame2011}. Assuming a constant rms noise across the cube, a voxel is retained only if it contains significant emission, and this emission is also present in its immediate spatial and spectral neighbourhood. Specifically, for each voxel at spatial position $(i,j)$ and velocity channel $k$, we required that: (i) the voxel itself and all eight spatial neighbours exceed the detection threshold $T \geq 1\sigma$, and (ii) the same condition is satisfied for the corresponding pixels in the neighbouring velocity channels $k-1$ and $k+1$. Only voxels satisfying these constraints are included in each moment-0 map. 

\begin{figure*}[h]
\centering
\includegraphics[width=0.80\hsize]{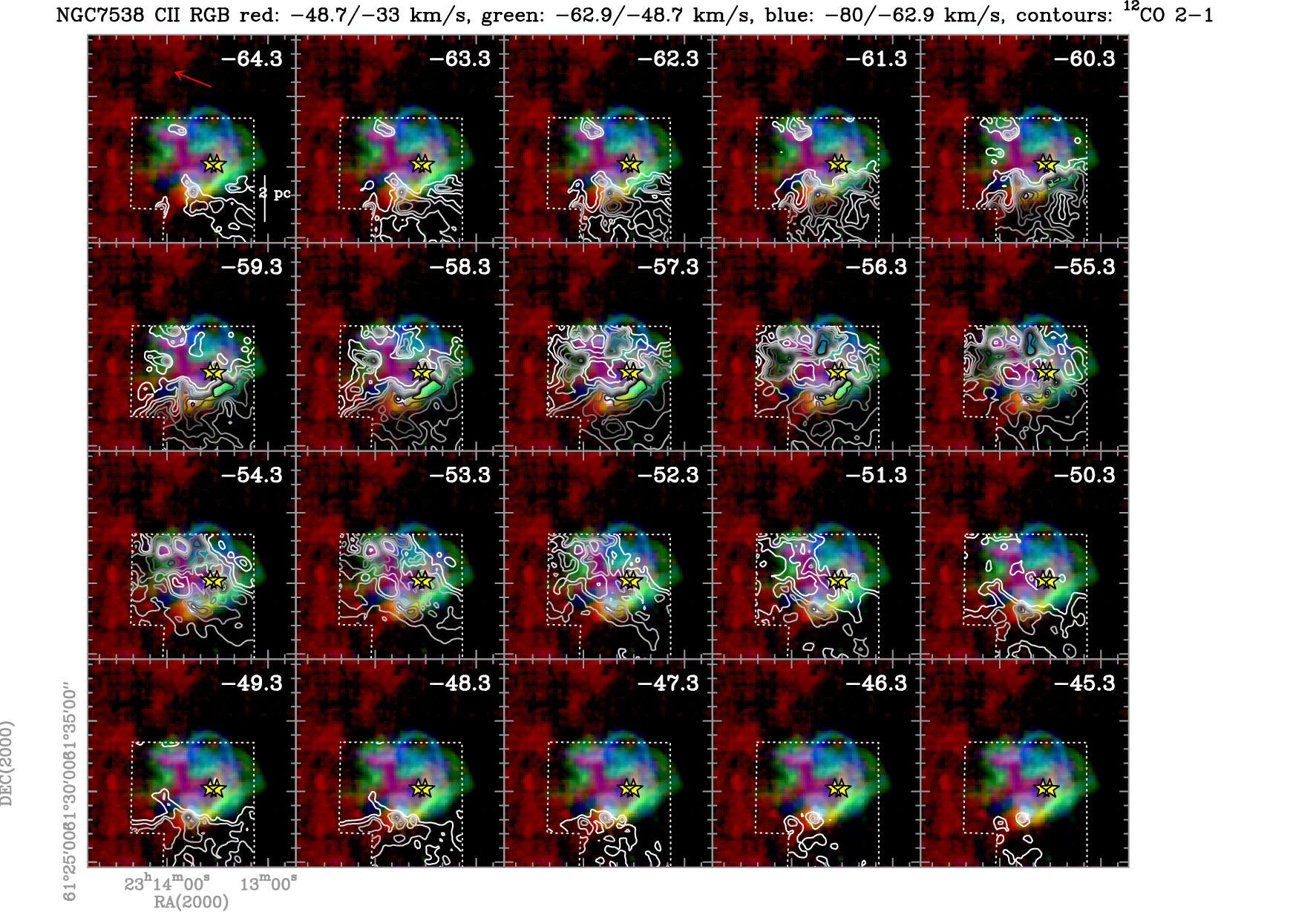}
\caption{
Combined RGB plot and channel map and \textit{Herschel} column density map for NGC7538. Each panel displays the distinct velocity ranges (blue and red for the high-velocity blue- and redshifted emission, green for the bulk emission) as an RGB plot with contours of $^{12}$CO(2-1) emission overlaid. The contour levels are 4 to 38 K km s$^{-1}$ in steps of 4 K km s$^{-1}$. The corresponding CO velocity is given in the upper-right corner. The two exciting stars are indicated with yellow star symbols. A very prominent example of high-velocity redshifted \CII\ gas not associated with an expanding bubble is indicated with a red arrow. 
}
\label{fig:rgb1}
\end{figure*}

\begin{figure*}
\centering
\includegraphics[width=0.76\hsize]{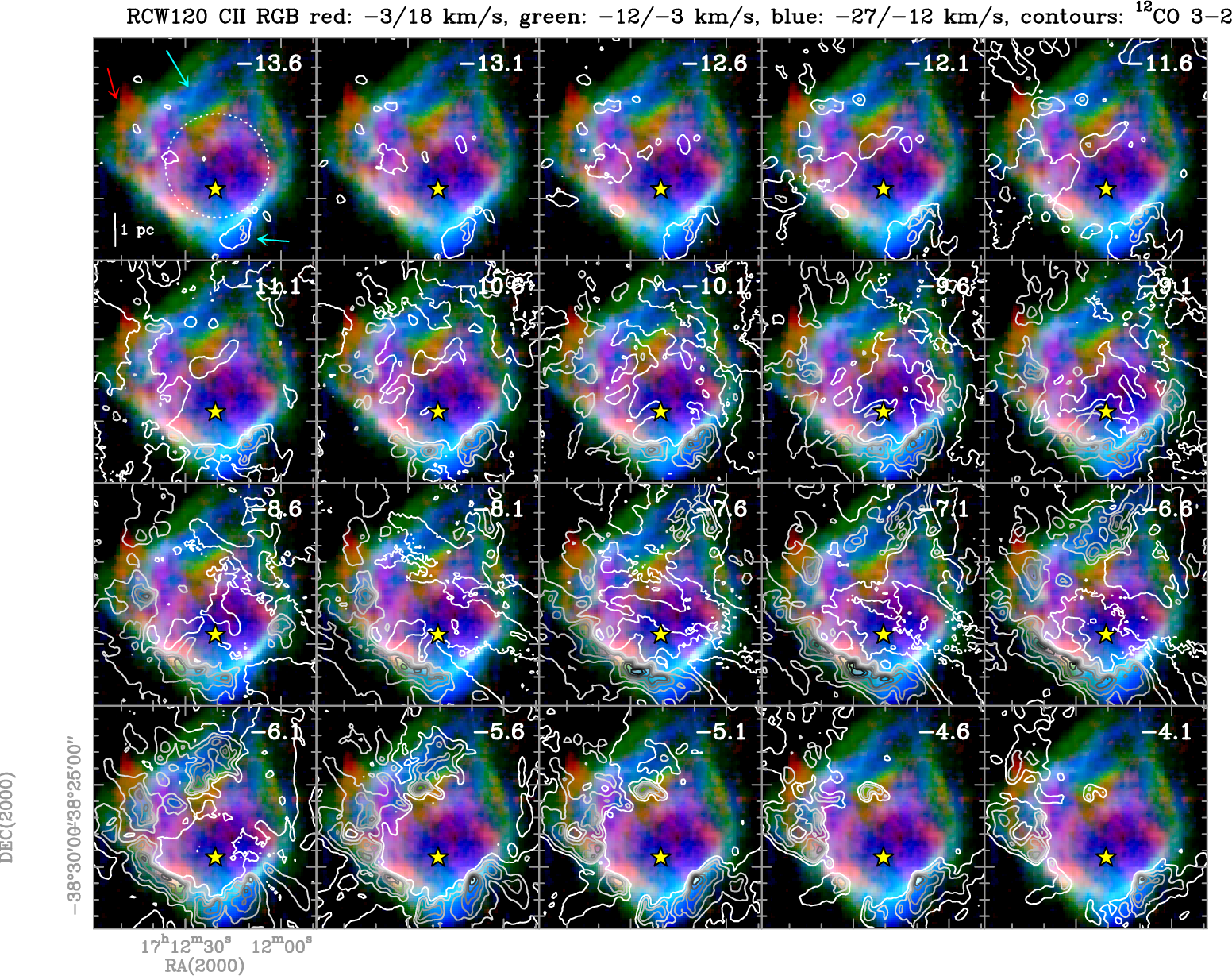}
\centering
\makebox[\linewidth][r]{
\includegraphics[width=0.84\hsize]{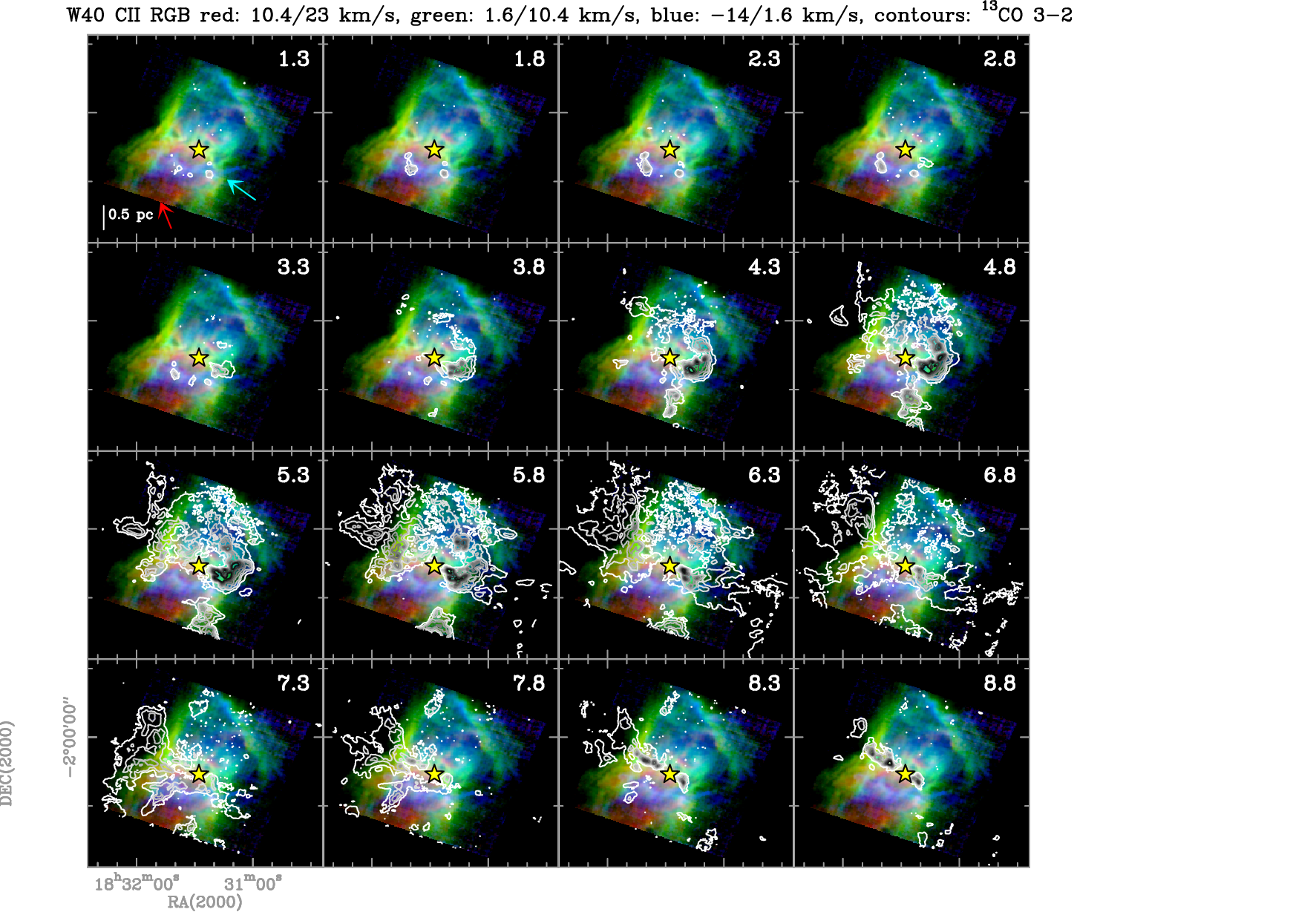}}
\caption{
Combined RGB plot and channel maps for RCW120 and W40. Each panel displays the distinct velocity ranges (blue and red for the high-velocity blue- and redshifted emission, green for the bulk emission) as an RGB plot with contours of $^{12}$CO(3-2) (RCW120) and  $^{13}$CO(3-2) (W40) emission overlaid. The contour levels are 4 to 39 K km s$^{-1}$ in steps of 5 K km s$^{-1}$ for RCW120 and 2 to 26 K km s$^{-1}$ in steps of 4 K km s$^{-1}$ for W40. The corresponding CO velocity is given in the upper-right corner. In the first RCW120 panel, the expanding \CII\ bubble is approximated by a dotted circle. The exciting stars are  indicated with yellow star symbols. Examples of high-velocity blue- and redshifted \CII\ gas not associated with the expanding bubble are indicated with blue and red arrows, respectively.
}
\label{fig:rgb2}
\end{figure*}

\begin{figure*}
\centering
\includegraphics[width=0.75\hsize]{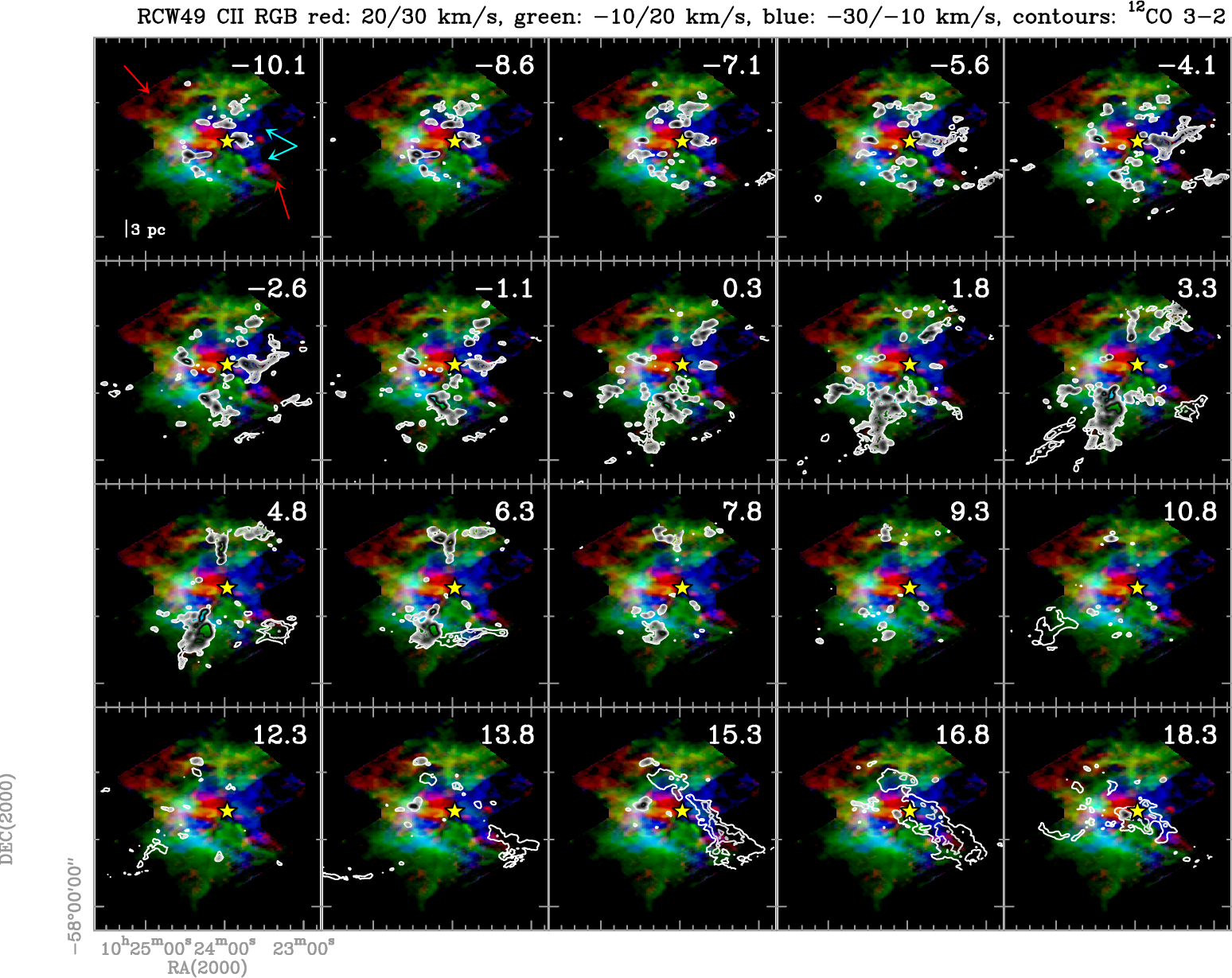}
\includegraphics[width=0.75\hsize]{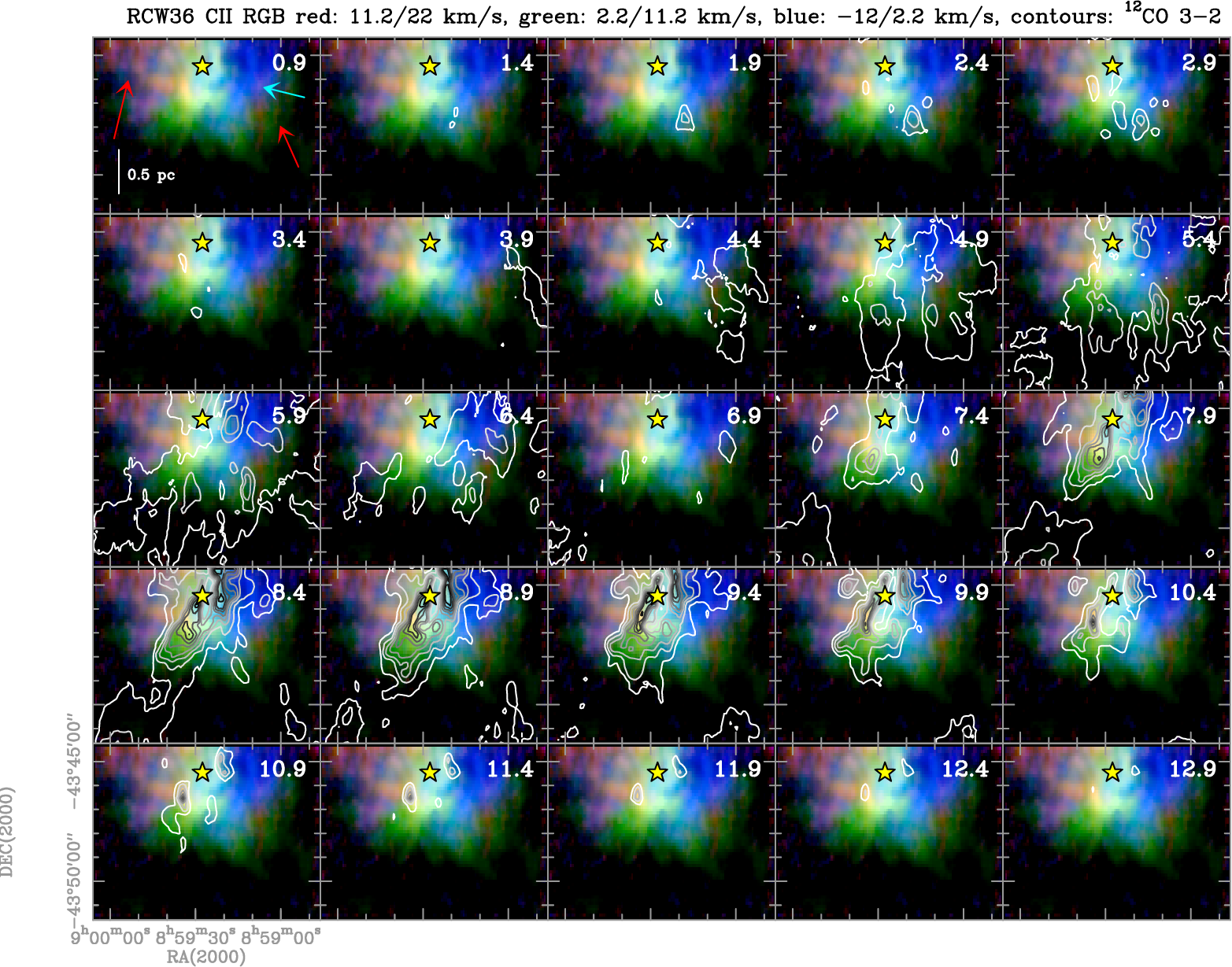}
\caption{
Combined RGB plot and channel maps for RCW49 and RCW36. Each panel displays the distinct velocity ranges (blue and red for the high-velocity blue- and redshifted emission, green for the bulk emission) as an RGB plot with contours of $^{12}$CO(3-2) emission overlaid. The contour levels are 3 to 23 K km s$^{-1}$ in steps of 2.5 K km s$^{-1}$ for RCW49 and 5 to 53 K km s$^{-1}$ in steps of 6 K km s$^{-1}$ for RCW36. The corresponding CO velocity is given in the upper-right corner. The exciting star or star cluster is indicated with a yellow star symbol. Examples of high-velocity blue- and redshifted \CII\ gas not associated with the expanding bubble are indicated with blue and red arrows, respectively. 
}
\label{fig:rgb3}
\end{figure*}

\begin{figure*}
\centering
\includegraphics[width=0.80\hsize]{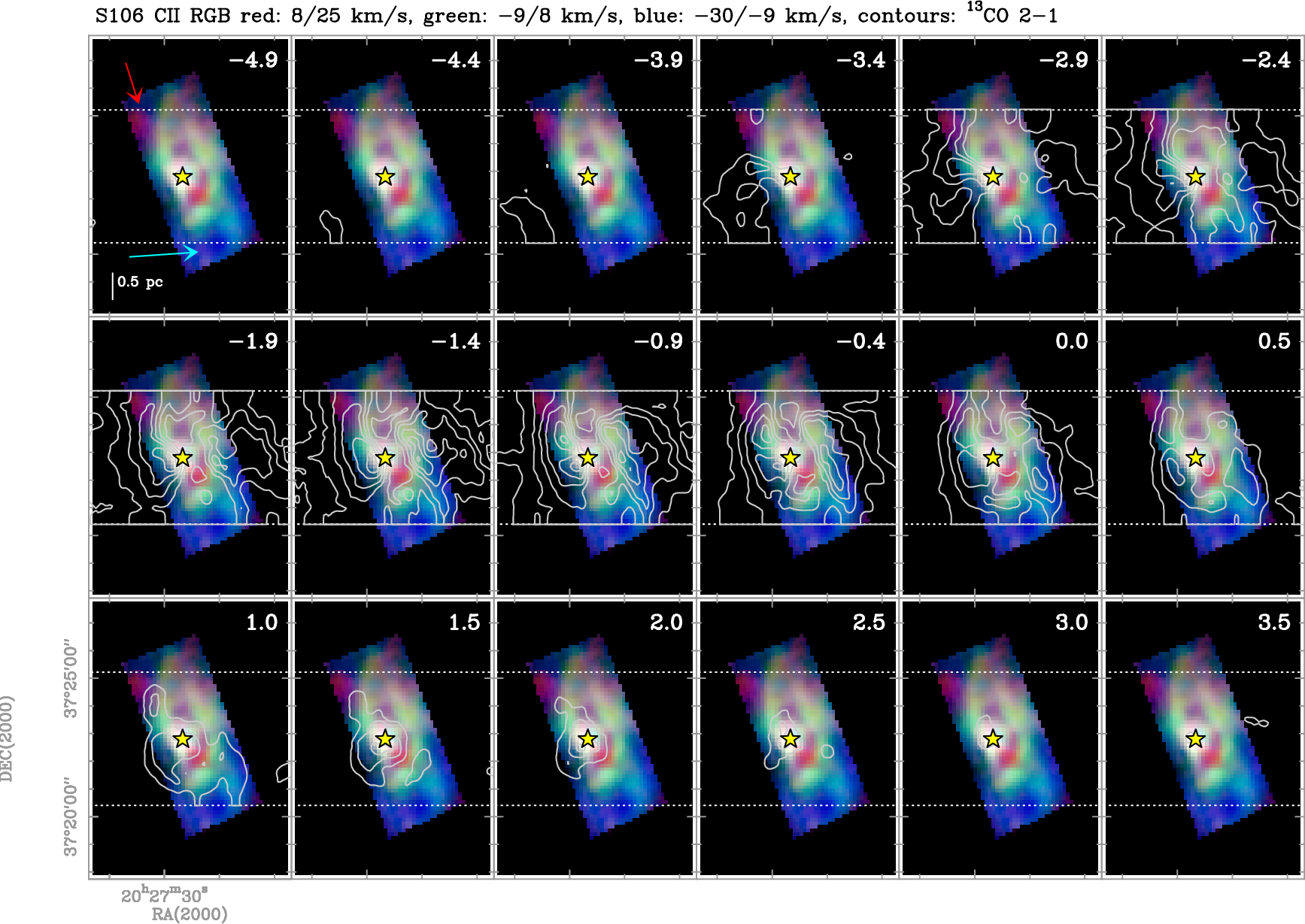}
\includegraphics[width=0.80\hsize]{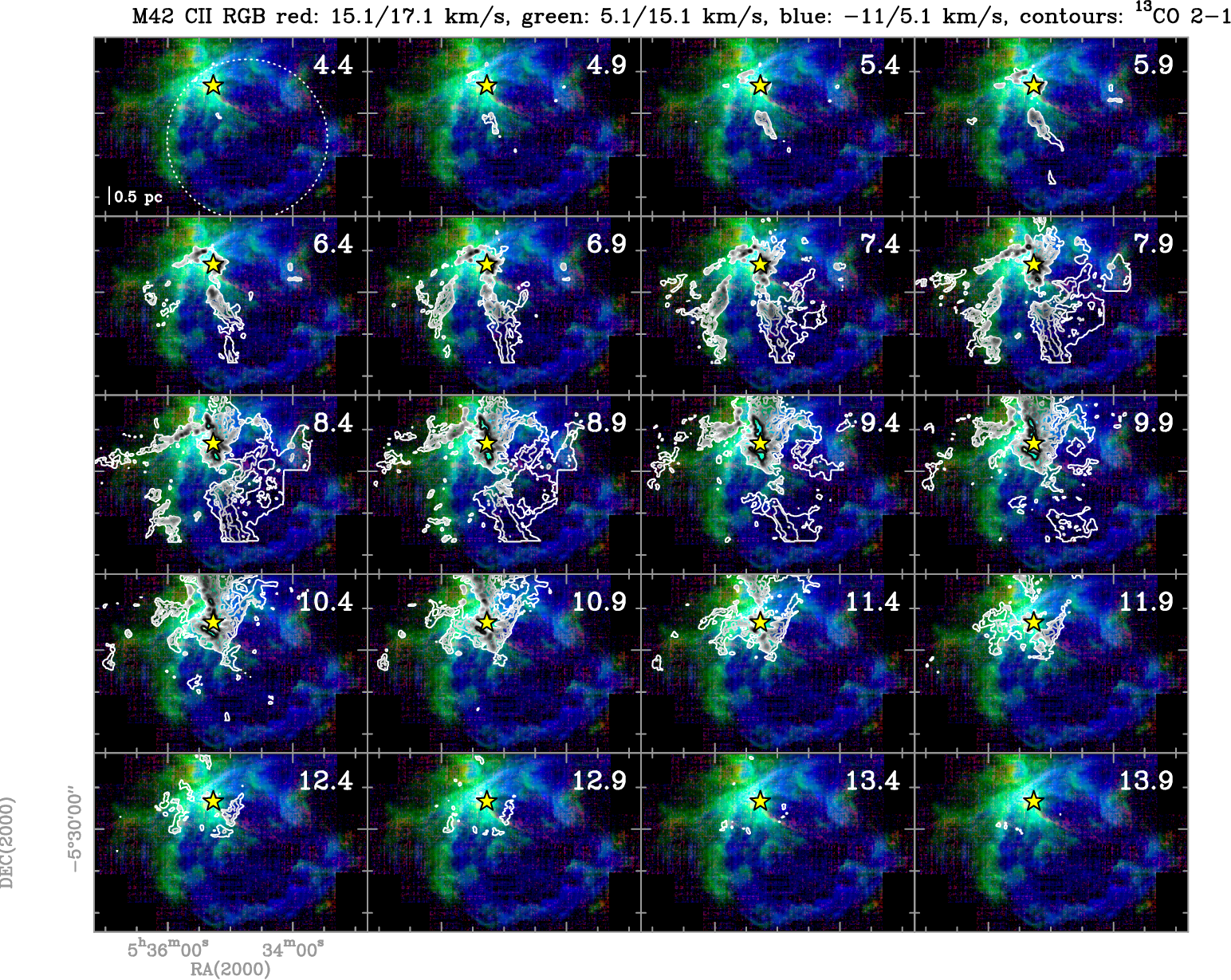}
\caption{
Combined RGB plot and channel map for S106 and M42.  In the first M42 panel, the expanding \CII\ bubble is approximated by a dotted circle. Each panel displays the distinct velocity ranges (blue and red for the high-velocity blue- and redshifted emission, green for the bulk emission) as an RGB plot with contours of $^{13}$CO(2-1) emission overlaid. The contour levels are 3 to 17 K km s$^{-1}$ in steps of 2 K km s$^{-1}$ for S106 and 3 to 20.5 in steps of 3.5 K km s$^{-1}$for M42. The corresponding CO velocity is given in the upper-right corner. The exciting star is indicated with yellow star symbols. 
}
\label{fig:rgb4}
\end{figure*}

\begin{figure*}
\centering
\includegraphics[width=0.77\hsize]{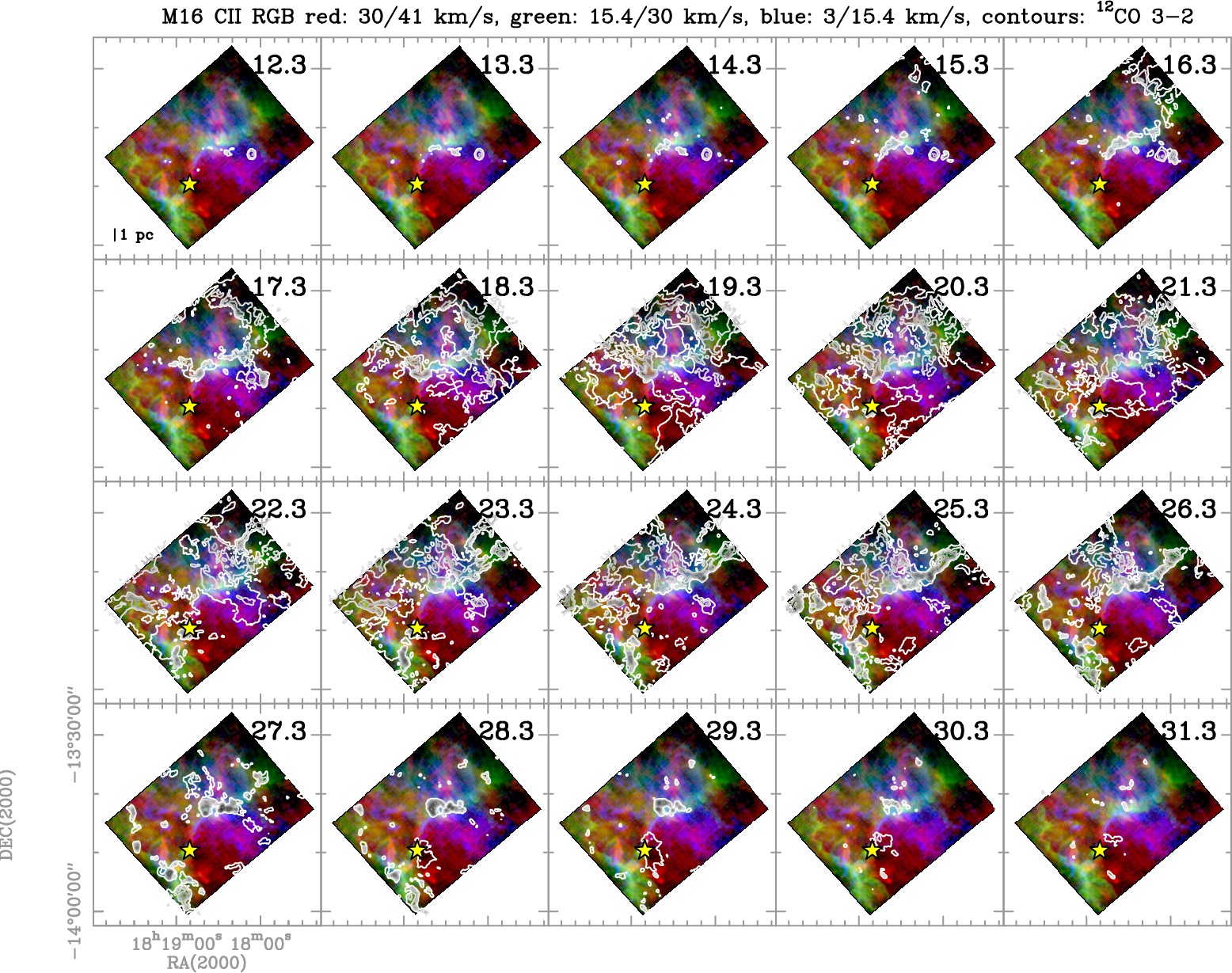}
\includegraphics[width=0.77\hsize]{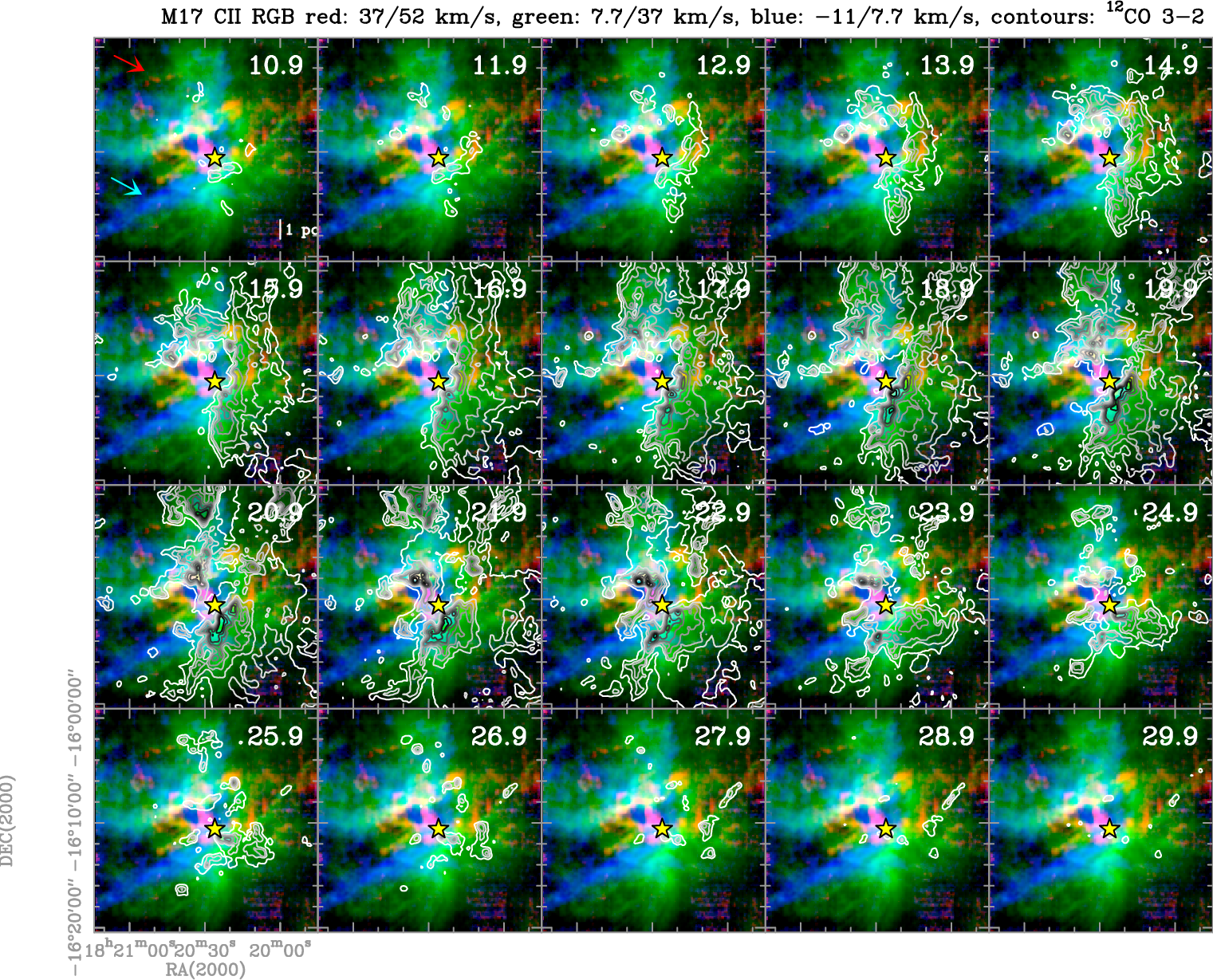}
\caption{
Combined RGB plot and channel maps for M16 and M17. Each panel displays the distinct velocity ranges (blue and red for the high-velocity blue- and redshifted emission, green for the bulk emission) as an RGB plot with contours of $^{12}$CO(3-2) emission overlaid. The contour levels are 5 to 68 K km s$^{-1}$ in steps of 9 K km s$^{-1}$ for M16 and 4 to 52 K km s$^{-1}$ in steps of 6 K km s$^{-1}$ for M17. The corresponding CO velocity is given in the upper-right corner. The exciting star clusters are indicated with a yellow star symbol. Examples of high-velocity blue- and redshifted \CII\ gas not associated with the expanding bubble are indicated with blue and red arrows, respectively. 
}
\label{fig:rgb5}
\end{figure*}

\newpage
\section{The dynamical timescale distributions}\label{sec:dyn_timescale_app}
Figure~\ref{fig:dyn_timescales_all} presents the obtained distribution of dynamical timescales associated with the high-velocity gas in M16, M17, M42, RCW 36, RCW 49, RCW120, NGC7538,  and W40. The distributions vary from region to region, but almost all the high-velocity gas has estimated dynamical timescales of a few hundred thousand years. Only in M16,  RCW 49, and RCW 79 (Fig. \ref{fig:dyn_age_hists}) are there small tails to the overall timescale distribution that result from a few pixels with dynamical timescales of about 1 million years.

\begin{figure*}[h]
    \centering
    \includegraphics[width=0.32\hsize]{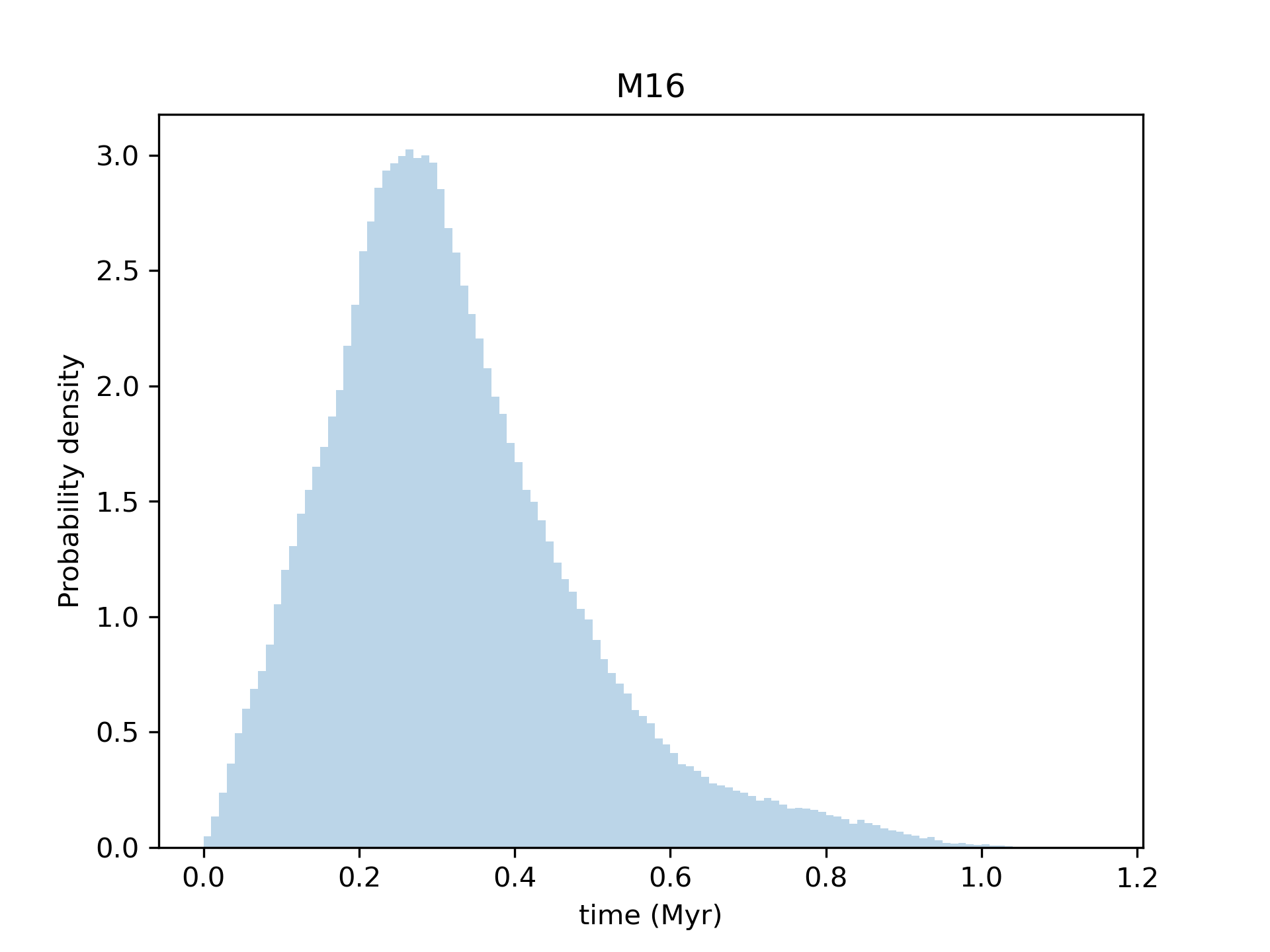}
    \includegraphics[width=0.32\hsize]{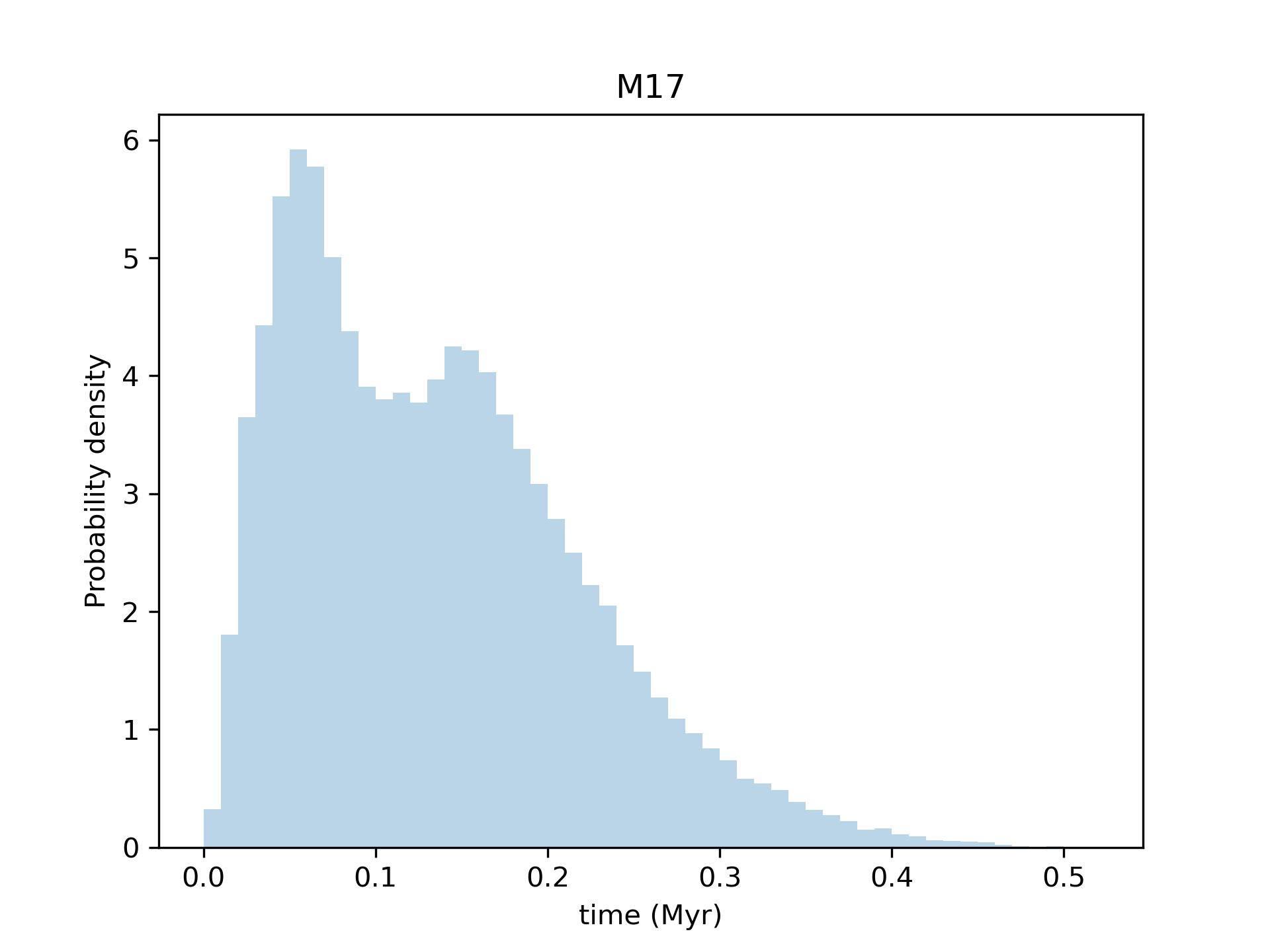}
    \includegraphics[width=0.32\hsize]{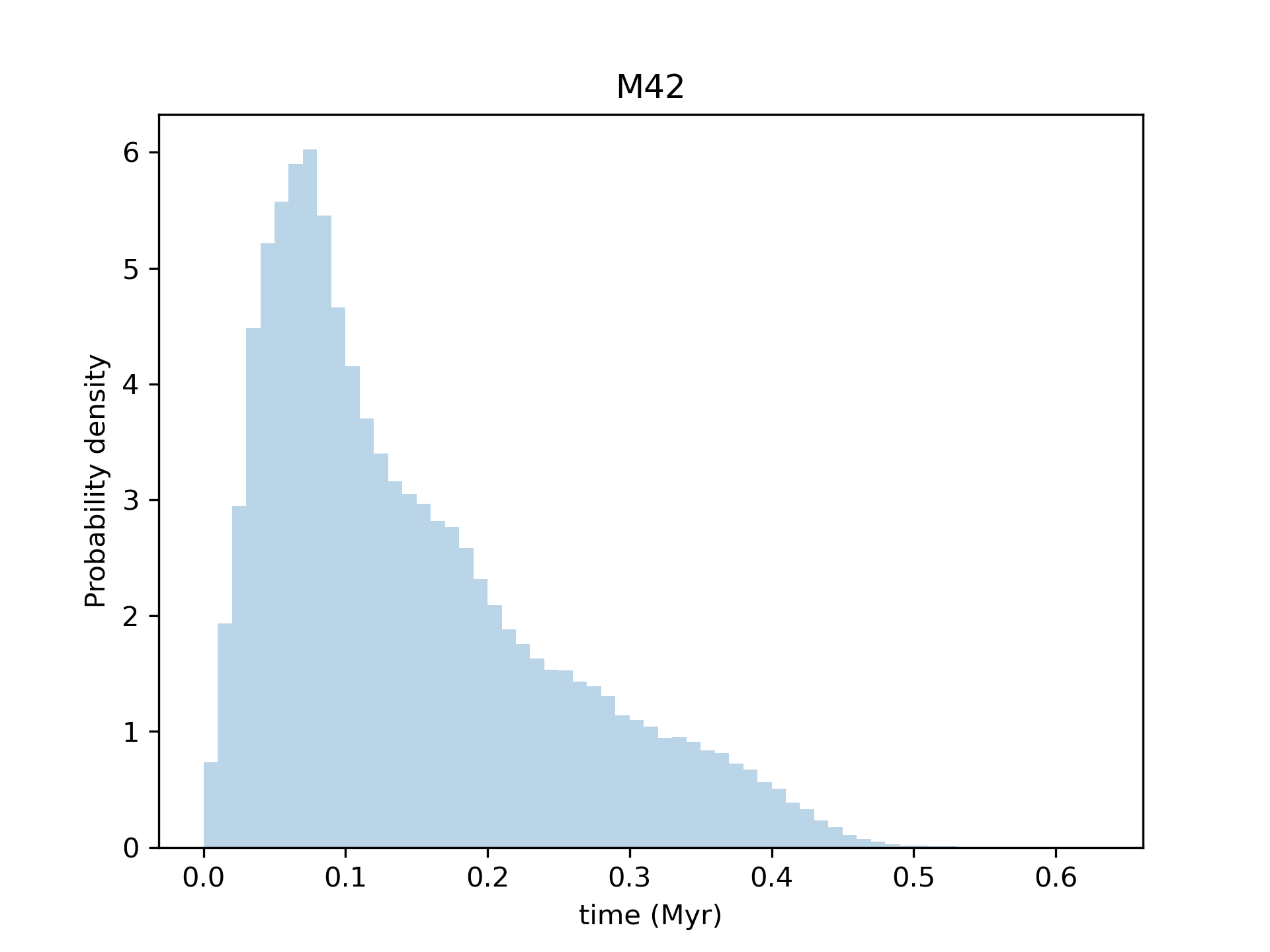}
    \includegraphics[width=0.32\hsize]{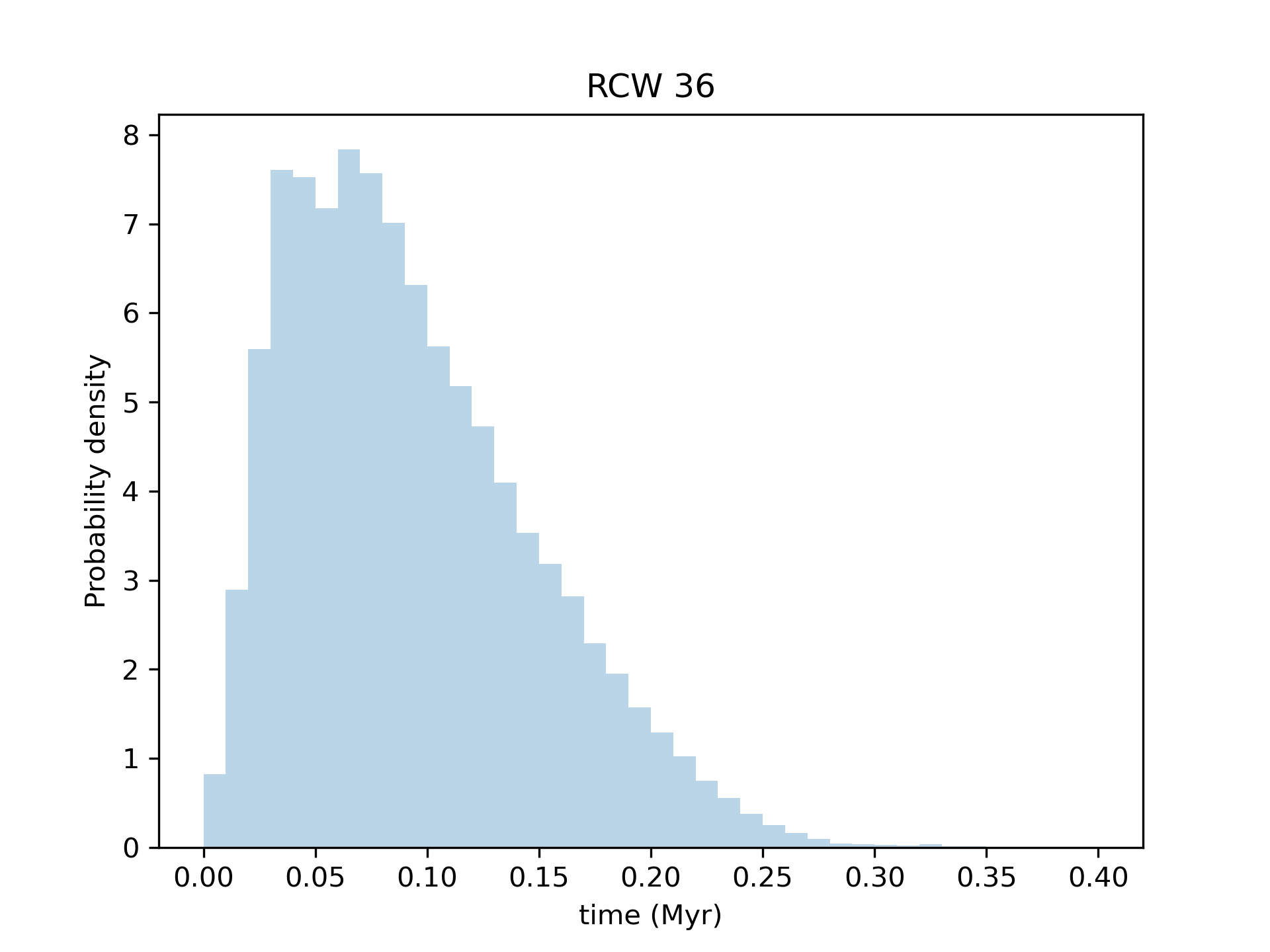}
    \includegraphics[width=0.32\hsize]{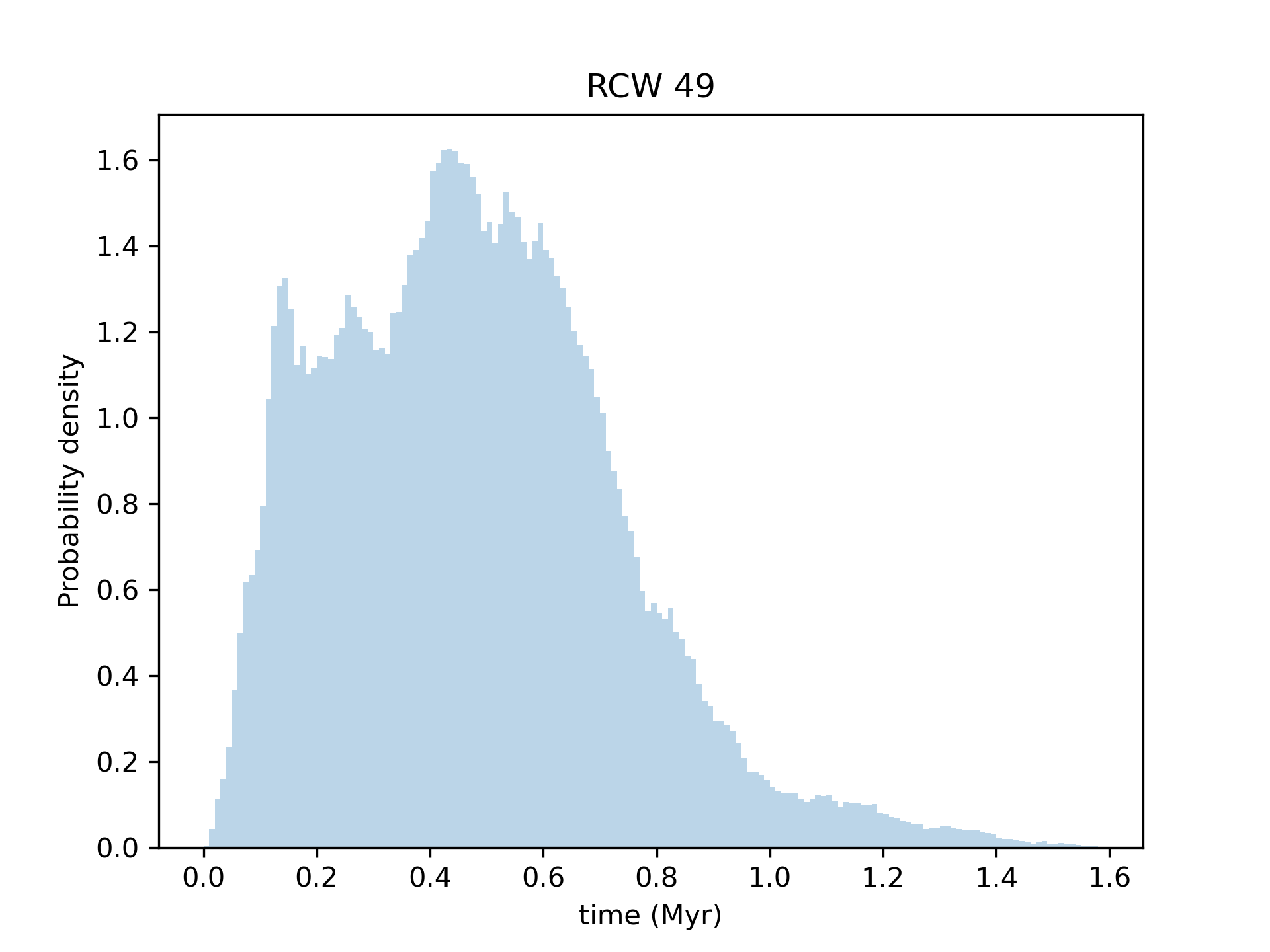}
    \includegraphics[width=0.32\hsize]{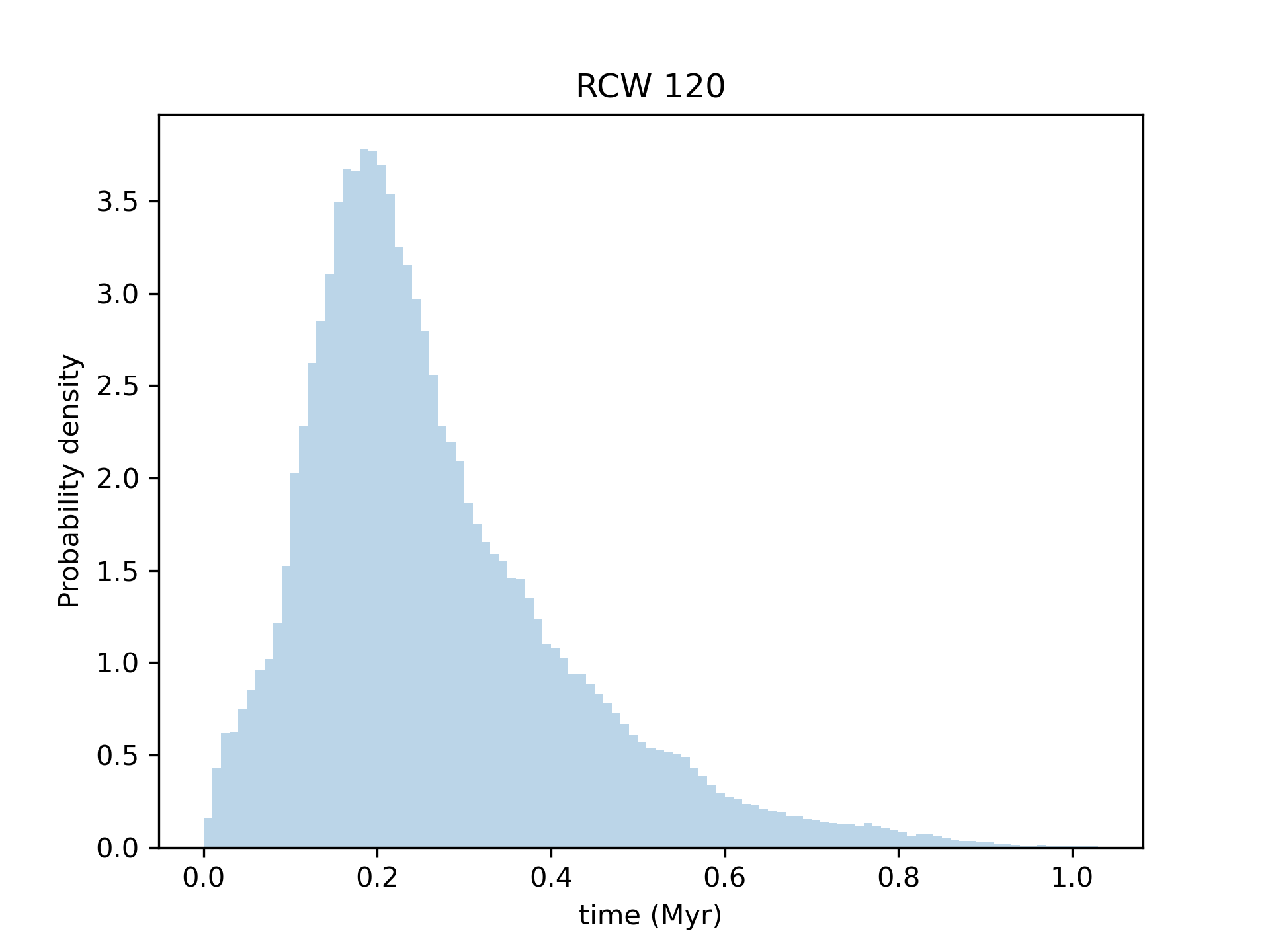}
    \includegraphics[width=0.32\hsize]{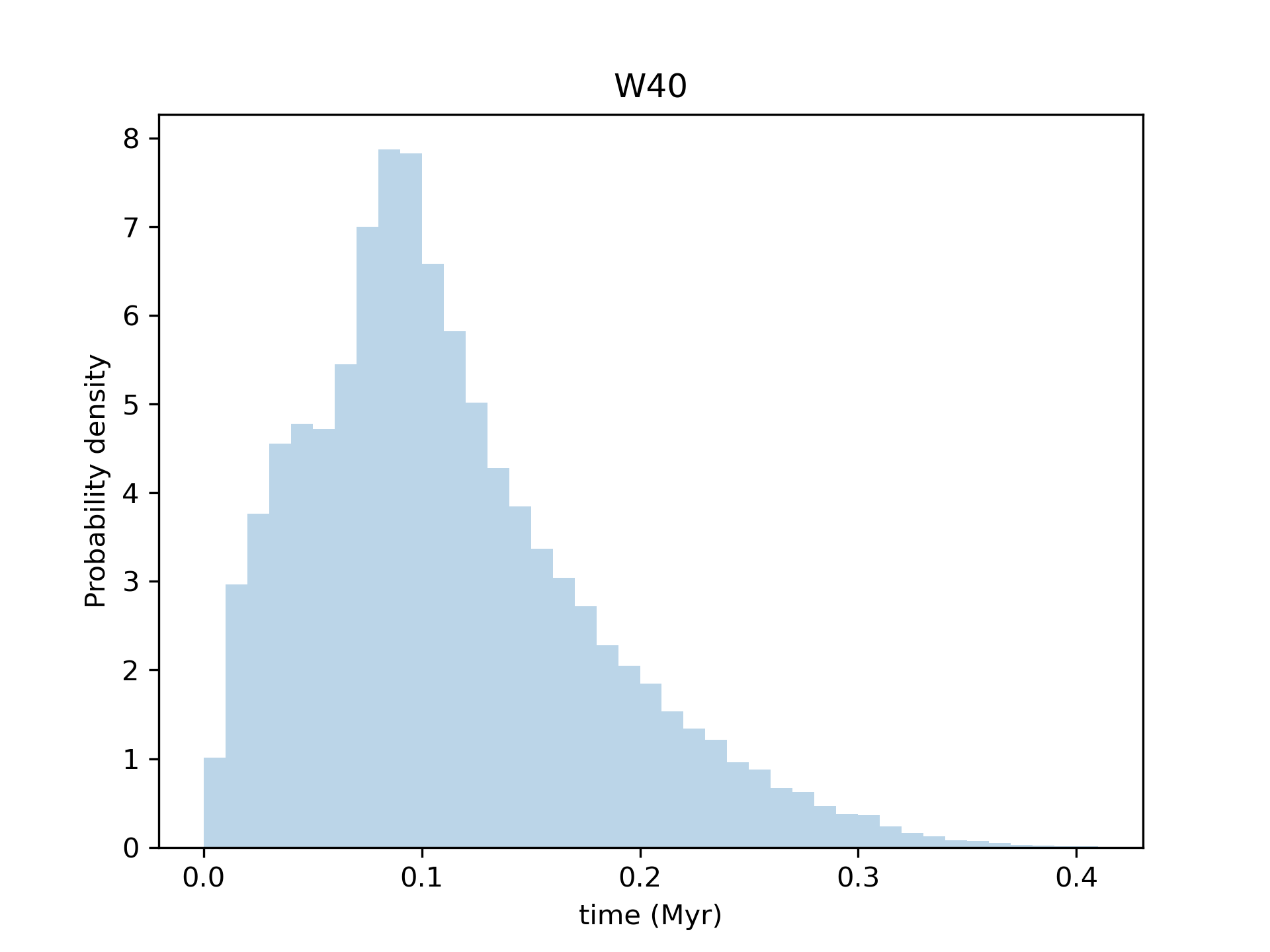}
    \includegraphics[width=0.32\hsize]{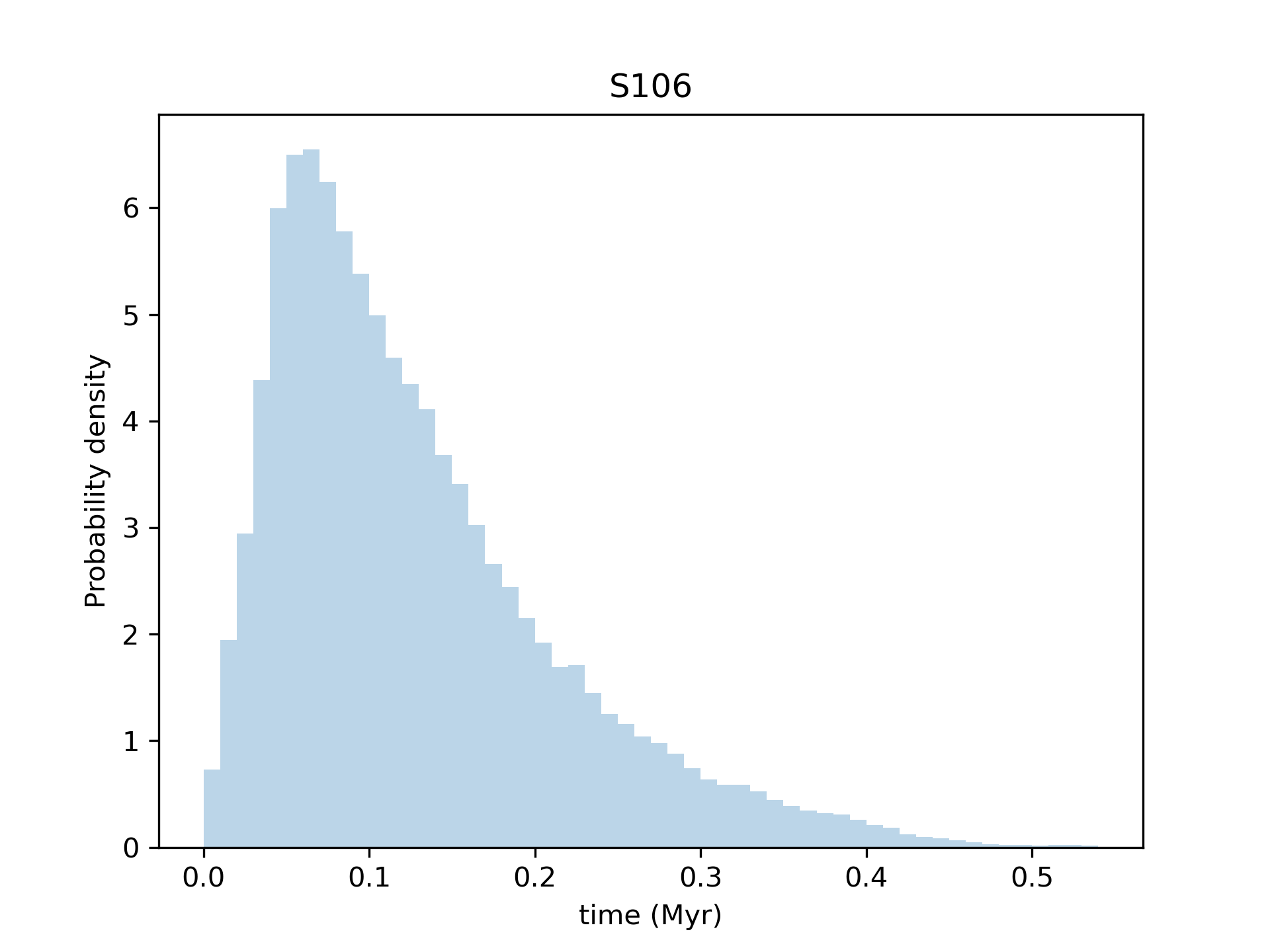}
    \includegraphics[width=0.32\hsize]{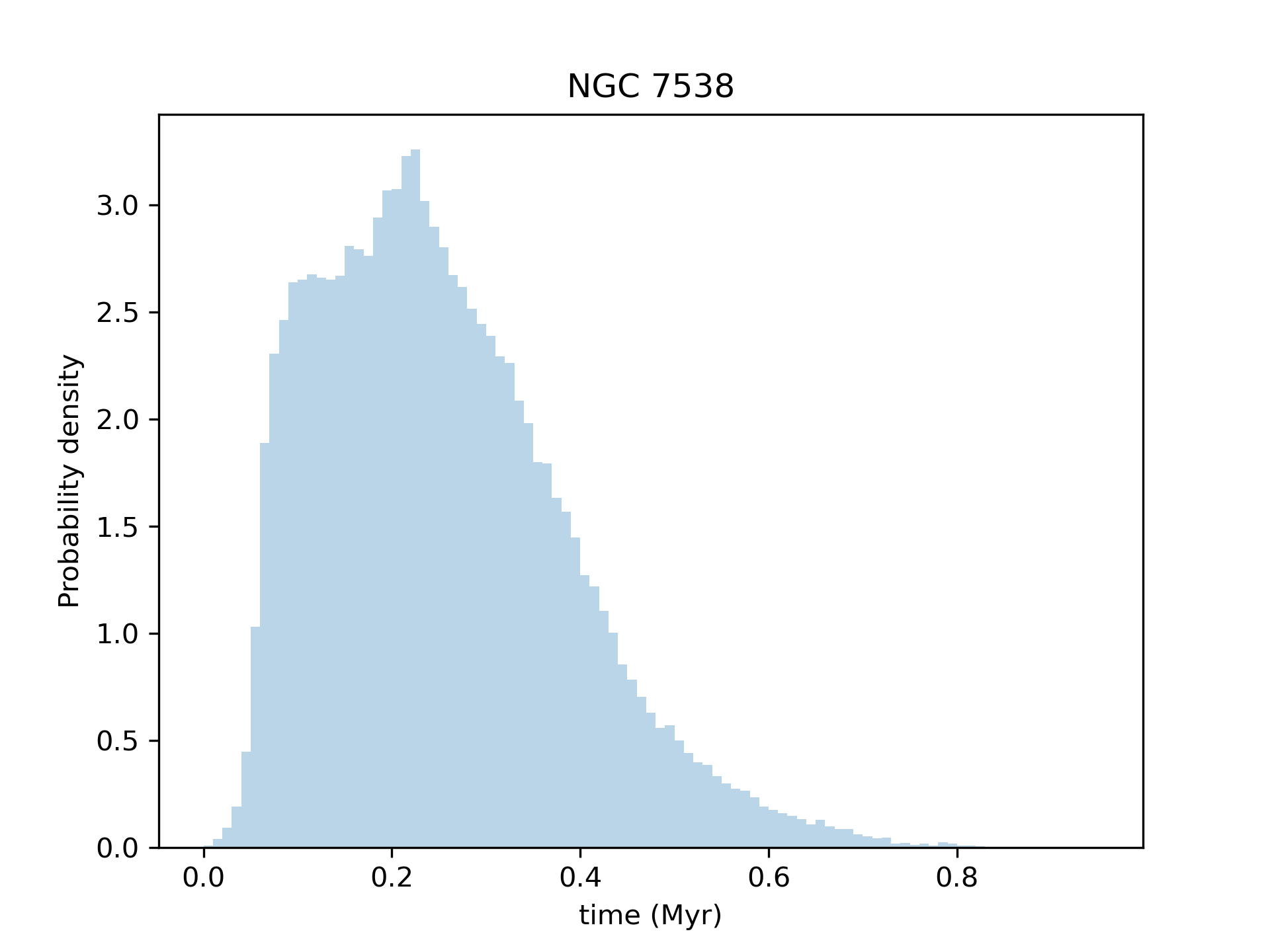}
    \caption{Distributions for the dynamical timescales associated with the high-velocity wings in M16, M17, M42, RCW36, RCW49, RCW120, W40, S106, and NGC7538.}
    \label{fig:dyn_timescales_all}
\end{figure*}

\end{appendix}

\end{document}